\documentclass{article}
\usepackage{times}
\usepackage[letterpaper, width=6.5in, height=9.0in]{geometry}
\usepackage[T1]{fontenc}
\usepackage{lmodern}

\usepackage{amsmath,amssymb}
\usepackage{booktabs}
\usepackage{graphicx}
\usepackage{float}
\usepackage[table]{xcolor}
\usepackage[hidelinks,bookmarks=false]{hyperref}
\newfloat{algorithm}{t}{loa}
\floatname{algorithm}{Algorithm}
\graphicspath{{figures/}}
\newcommand{\Description}[1]{}
\newcommand{\SIEVE}{\textsc{Sieve}}

\newcommand{\hitobj}{H_{\mathrm{obj}}}
\newcommand{\hitbyte}{H_{\mathrm{byte}}}

\title{Caching Policies based on Object Sizes}
\author{Ridwanul Hasan Tanvir and George Kesidis\\
\small Computer Science \& Engineering Dept, Electrical Engineering Dept\\
\small Pennsylvania State University\\
\small University Park, PA, 16802, USA\\
\small \texttt{\{rpt5409,gik2\}@psu.edu}}
\date{}

\begin{document}
\maketitle

\begin{abstract}
We evaluate eviction policies that consider the lengths of the objects being cached for 
different cache sizes (cost) and different workloads.
Workloads include
the Independent Reference Model and real-world production traces.
In particular, hybrid policies are considered for segmented caches.
A working-set approximation is developed for SIEVE as well as
its length-aware extension.
\end{abstract}


\section{Introduction}

Cache replacement is often described as choosing which resident object to evict.
For variable-length objects, a caching policy also
decides how much byte capacity remains for other objects. Admitting one large
object may require evicting hundreds of smaller objects, each of which may be very useful
for serving requests in the near future. This setting is common in web caches where
objects may range from small metadata records to videos, images, and software
artifacts~\cite{NygrenAkamai10,BergCacheLib20}.

The variation in object size is not confined to one cache or one workload.
Studies of large-scale Content Distribution Networks (CDNs) and in-memory caches report substantial diversity in
both object sizes and access patterns~\cite{BergerAdaptSize17,YangTwitter20,
BergCacheLib20}. This matters because replacement policies operate under a byte
constraint even when their state is organized around objects. AdaptSize, for
example, showed that size-aware admission can substantially change the object
hit rate of a CDN cache~\cite{BergerAdaptSize17}, while the LHD policy explicitly
accounts for the expected number of hits relative to the space consumed by an
object~\cite{BeckmannLHD18}. In both systems, object size directly affects
which objects the cache should keep.

Consider a cache that serves small API responses, thumbnails, HTML fragments,
small images, videos, and software images from the same byte budget. In one
production trace we consider, the median object is tens of kilobytes (kB), while the largest
object is gigabytes (GB). A single $2$GB object can occupy the space of roughly
fifty thousand $40$kB objects. If a size-oblivious policy admits and protects
that large object after one reference, it may evict many small objects that would
have served later requests, i.e., those evictions become backend cache misses. The
cache may still serve nearly the same byte volume, but it serves fewer requests.
The size-aware policies in this paper are designed to avoid this outcome: they
retain many small reusable objects unless a large object has shown
stronger evidence of current reuse.

The value of size awareness is workload dependent. It is most useful when the
objects competing for cache space are both popular and heterogeneous in size,
and when their aggregate footprint is larger than the cache budget. If the
currently popular objects all fit in the cache, or if the popular objects have nearly equal sizes, a
size-oblivious policy loses little by ignoring size. The difficult case is the
``middle'' one: popular objects overlap in time, compete for the same bytes, and
vary enough in size that protecting one large object can displace many smaller
objects with comparable current-reuse evidence.

There is, however, no universal rule that smaller objects should receive higher
priority. The appropriate treatment of size depends on what the cache is intended
to optimize. A request-oriented cache may benefit from retaining more objects
that are likely to be requested again, whereas a bandwidth-oriented cache may
place greater value on retaining large objects. A particular application
using a cache may also attach
different costs or importance to different objects. Thus, size is a signal used
by the replacement policy; it does not by itself define the objective. In this
paper we evaluate the policies from the cache's point of view using conventional
object and byte hit rates, rather than assuming an application-specific weighting
of objects. This said, we consider different cases of cache capacities as a signal for the
\textit{cost} of the cache, noting that cache users may have Service-Level Objectives
(SLOs, in terms of hit rates) which they trade off with capacity-dependent
cost as may be specified in
a Service-Level Agreement (SLA) of the cache.

An operator can state the design goal in several ways.
One objective is to minimize cache capacity while meeting a target hit rate:
$\min B$ subject to $H \ge H^\star$.
Another possible objective fixes the available byte budget and asks for the highest hit
rate:
  $\max H$ subject to $B \le B^\star$.
A third possible objective combines performance and resource cost directly:
 $ \max ~ H-\lambda B$,
where \(\lambda>0\) is the operator's price for resident capacity. 
Generally non-convex utility
formulations of these kinds have been used for cache design and cache
policies~\cite{DehghanUtility16,NegliaCM18}. The symbol
\(H\) can denote object hit rate, byte hit rate, or an application-defined
utility-weighted hit rate. Our mechanisms instantiate the object-hit-oriented
choice: all other reuse evidence being equal, a shorter object receives more
retention priority because the same byte budget can hold more such objects. A
different deployment could reverse or reshape this preference. For example, a
cache for large media objects, software images, or costly-to-restore data may
prefer to protect larger objects under the same framework.

This paper focuses on object hit rate. Object hit rate measures the fraction of
requests served from cache, and it is the natural metric when backend request
load is the bottleneck. Byte hit rate is also important, but it measures a
different objective. A cache that keeps many small popular objects can serve many
more requests while moving about the same number of bytes. We therefore optimize
for object hit rate and check byte hit rate separately. We also report policy computational
overhead, since replacement logic consumes CPU and metadata bandwidth and can limit
cache throughput~\cite{DongLockFree25,QiuHitRatioThroughput26}, even though
its cost may be partly hidden by (pipelined with) the I/O overhead needed to fetch or move large objects.

Prior size-aware designs intervene at different points in cache management.
AdaptSize uses object size at admission time and adapts the admission rule as the
workload changes~\cite{BergerAdaptSize17}. LHD instead estimates the expected
hits obtained per unit of cache space and uses that estimate to guide
replacement~\cite{BeckmannLHD18}. Both approaches demonstrate that size can be
useful when the objective is request hit rate, but they introduce mechanisms
specifically designed for size-aware admission or ranking. We ask a
narrower question: how much of this benefit can be obtained by adding a small,
causal, size-dependent decision to replacement policies that already have their
own notions of recency, reuse, admission, and promotion?

We start with \SIEVE{}, a FIFO-like policy that uses one
visited bit and an eviction hand~\cite{ZYY24}. We then apply the same size signal
inside three common blended families: 2Q, ARC, and S3-FIFO. In every case, the
policy sees only past requests and current object sizes. It does not use future
reuse, offline labels, or an oracle.

The evaluation follows the same point of view. Rather than selecting one cache
size and declaring a winner, we compare policies across byte budgets and ask how
object hit rate changes with the amount of cache memory allocated. This exposes
the performance--capacity tradeoff that a cache operator faces when provisioning
memory~\cite{SundarrajanCoNEXT17}. The synthetic
traces then isolate one part of that tradeoff by holding the request process
fixed while increasing the coefficient of variation of object sizes.

\SIEVE{} is a natural starting point because its replacement state is small. 
A hit records reuse by setting a visited bit, while
the ordering of resident objects is left unchanged until the eviction hand
examines them~\cite{ZYY24}. The same work also showed that \SIEVE{} can serve as
a replacement primitive inside policies such as 2Q and ARC, rather than only as
a standalone cache policy. Canonical \SIEVE{}, however, gives the same
replacement state to objects of different sizes. This leaves a direct question
for variable-length caches: whether size can be incorporated without giving up
the simple state and deferred eviction decisions that make \SIEVE{} attractive.

A deployed cache should not need a new cache architecture to obtain a basic
size signal. We therefore keep the mechanisms close to their parent policies,
which also lets us measure the effect of the size signal directly.

A second question is whether the resulting behavior admits a useful analytical
description. Characteristic-time and working-set approximations have provided a
tractable way to reason about cache occupancy and hit probability for policies
such as LRU under the Independent Reference Model (IRM)~\cite{CTZ02,
FrickerRoberts12,MartinaGaretto14,GastVanHoudt15,GarettoLM16,FerragutRP16}. 
\SIEVE{} does not maintain an LRU stack:
objects remain in insertion order, and eviction is determined by successive
examinations of the hand. In particular, the first examination after admission
is different from subsequent examinations of an object that has already survived
one examination. We therefore examine whether these two stages can be represented by
separate characteristic times, and whether the resulting approximation can be
adapted from an object-count constraint to the byte-capacity constraint of a
variable-length cache.

This paper makes the following contributions.
\begin{itemize}
  \item We describe two new causal size-aware variants to \SIEVE{}
  (Section~\ref{sec:sieve-variants}). Mean-Threshold grants
  extra protection below the running mean object size, while Dual-Threshold
  separates small, medium, and large objects using the running mean. On the
  production traces, Dual-Threshold increases object hit rate by up to $27.7\%$
  over canonical \SIEVE{}.
  \item We evaluate size-aware policies across production and synthetic traces,
  including LRU-MIN (reported as LRU\textsuperscript{!}), Dual-Threshold
  \SIEVE{} (reported as \SIEVE{}\textsuperscript{!}), and new size-aware segmented
  versions of 2Q, ARC, and S3-FIFO (Section~\ref{sec:segmented-size-aware}). The hybridized \SIEVE{}-based policies give the
  best object hit rate most often: 2Q-SIEVE\textsuperscript{!} on the META
  traces and ARC-SIEVE\textsuperscript{!} on the synthetic traces.
  \item We compare cache-eviction policies in terms of hit rate versus
  cache-memory size rather than at a single fixed cache size. This is the
  performance--cost tradeoff seen from a user's point of view.
  \item We develop a new working-set approximation for \SIEVE{} under the
  IRM (Section~\ref{sec:wsa} and
  Appendix~\ref{app:wsa-details}). The model uses two examination times and handles
  variable-length objects by replacing the unit-capacity equation with a
  byte-weighted one.
\end{itemize}

This paper is organized as follows. Section~\ref{sec:background} introduces the
cache model, size-aware replacement mechanisms, and the policies evaluated in
the paper. Section~\ref{sec:methodology} describes the traces, metrics, and
experimental setup. Sections~\ref{sec:sieve-results}--\ref{sec:overall-results}
present the empirical results, Section~\ref{sec:wsa} develops the working-set
approximation, and Sections~\ref{sec:discussion} and~\ref{sec:conclusion}
discuss implications and conclude.


\section{Background and Size-Aware Cache Replacement}
\label{sec:background}

\subsection{Cache Model and Evaluation Objectives}

A content cache stores objects so that subsequent requests can be served without
accessing a slower backend, such as a remote origin server, a lower cache tier, or
persistent storage. Modern web and storage caches serve workloads with many
distinct objects, skewed request popularities, and substantial variation in object
sizes~\cite{BergCacheLib20,YangTwitter20,BergerAdaptSize17}. A cache therefore
has to decide not only which objects have sufficient evidence of reuse, but also
how to allocate a finite byte budget among objects that may occupy very
different amounts of memory.

Let the resident cache capacity be \(B\) bytes. A request at time \(t\) is for
object \(x_t\) whose size is \(L_{x_t}\) bytes. A request is a \emph{hit} if
\(x_t\) is resident when the request arrives and a \emph{miss} otherwise. On a
miss, the requested object may be admitted to the cache; if there is
insufficient free space, the replacement policy evicts one or more resident
objects. Objects with \(L_{x_t}>B\) cannot be admitted and are treated as
forced misses.

We distinguish resident objects from policy metadata. A \emph{resident entry}
contains the cached payload together with the metadata needed by the
replacement policy. A \emph{ghost entry} records the identity of a recently
evicted object but does not retain its payload. A request to a ghost entry is
therefore still a miss. The ghost hit nevertheless provides causal evidence
that the evicted object has returned, and several segmented policies use this
information when making subsequent admission, promotion, or partitioning
decisions~\cite{JohnsonShasha94,MegiddoModha03,YZQ23}. As in the original
definitions of these policies, ghost memory is relatively small and is typically
not counted toward the resident byte capacity \(B\).

We evaluate two conventional cache metrics. Let \(I_t=1\) when request at time \(t\)
is a hit and \(I_t=0\) otherwise. Summing over request times, the object and byte hit rates 
respectively are
\begin{equation}
\hitobj =
\left. \sum_t I_t\middle/\sum_t 1\right.,
\qquad
\hitbyte =
\left. \sum_t I_t L_{x_t}\middle/ \sum_t L_{x_t}\right. .
\label{eq:hit-rates}
\end{equation}
Object hit rate gives equal weight to requests, whereas byte hit rate weights a
hit by the number of bytes served. These objectives need not favor the same
resident set. In particular, retaining several frequently requested small
objects may increase object hit rate without producing a corresponding increase
in byte hit rate.

Object size does not, by itself, determine the value of an object. An
application may prefer smaller objects, larger objects, or may assign
application-specific values to individual objects. Such preferences could be
represented by a weighted hit-rate objective or by a different size-dependent
retention rule. We do not assume a universal ordering of objects by size.
Instead, this paper uses conventional object hit rate as its primary metric and
evaluates byte hit rate separately. The size-aware mechanisms studied below
instantiate one particular retention rule---additional protection for objects
below an online size threshold---and evaluate its consequences under these
cache-side metrics.

\subsection{Replacement Policies and Object Size}

Classical cache-replacement policies infer future usefulness from recency,
frequency, or combinations of the two. LRU orders objects by their most recent
access, while policies such as LRU-K, 2Q, ARC, CAR, LIRS, and CLOCK-Pro retain
additional history or organize objects into multiple regions to distinguish
short-term recency from stronger evidence of reuse~\cite{ONeilLRUK93,
JohnsonShasha94,MegiddoModha03,BansalCAR04,JiangLIRS02,JiangClockPro05}.
More recent policies have shown that substantially simpler FIFO-like structures
can also obtain strong hit-rate performance. S3-FIFO uses FIFO queues,
lightweight counters, and ghost history~\cite{YZQ23}, while \SIEVE{} combines
insertion order with one visited bit and an eviction hand~\cite{ZYY24}. Such
CLOCK-like policies can show performance cliffs, where added capacity has
little effect until a workload-dependent threshold is
crossed~\cite{CidonCliffhanger16}; Wang et al. analyze a randomized SIEVE/CLOCK
variant with multiple access bits per object that addresses this
behavior~\cite{WangSieveClock26}.

Variable-length objects change the replacement problem because resident objects
consume unequal portions of the cache. For equal-sized objects, an eviction
releases one cache slot. For variable-length objects, the number of objects that
must be removed depends on the sizes of both the incoming and resident objects.
The offline optimum is therefore no longer characterized by the usual
equal-size form of Belady's rule, and optimal caching with variable object
sizes is substantially more difficult~\cite{Belady66,BergerFOO18}. Even the
equal-size optimum requires dedicated tools to compute on large
traces~\cite{HanScaleOPT24}.

Prior work incorporates size at several different points in cache management.
GreedyDual-Size combines object size with retrieval cost in the replacement
priority~\cite{CaoIrani97,Young02}. AdaptSize uses
size to control admission while retaining LRU as the underlying eviction
policy~\cite{BergerAdaptSize17}. LHD estimates the expected utility of an
object relative to the cache space and time that it occupies~\cite{BeckmannLHD18}.
Other approaches modify LRU itself so that victim selection depends jointly on
recency, size, or reference count~\cite{WilliamsLRUMIN96,AggarwalSALRU99,
ChengLRUSP00}. These approaches show that size helps, but each adds it through a
different mechanism. Learning-based policies take another route and train a
model online to choose eviction victims: LRB approximates Belady's rule for CDN
caches~\cite{SongLRB20}, and GL-Cache learns eviction decisions for groups of
similar objects~\cite{YangGLCache23}. A separate line of work
treats caching as online learning and gives regret guarantees against
adversarial request sequences~\cite{BhattacharjeeRegret20,
LiOnlineCachingNetworks21,MhaisenOptimistic22}. Our study is complementary: it
evaluates simple eviction rules empirically under a byte budget.

We ask whether size can be introduced into a simple
replacement primitive without replacing the surrounding cache architecture.
The mechanisms are causal: decisions use only object sizes and request history
observed up to the current time. They do not use future requests, oracle reuse
distances, or offline phase labels. This allows the effect of the size signal to
be studied separately from changes to admission, ghost management, or
promotion logic.

\subsection{Canonical \SIEVE{}}
\label{sec:canonical-sieve}

By canonical \SIEVE{} we mean the original, size-agnostic \SIEVE{}
policy~\cite{ZYY24}, which gives every object the same replacement state
regardless of its size. It keeps resident objects in insertion order with one
visited bit per object. A hit only sets the bit. When space is needed, an
eviction hand scans from older to newer objects, clears the bits that are set,
and evicts the first object whose bit is zero. The hit path is therefore constant
time, and the consequence of a hit is decided only when the hand later examines
the object; the size-aware variants below change retention at this point.
Appendix~\ref{app:canonical-sieve} gives the full description and pseudocode.
Also see Appendix~\ref{app:wsa-details}.

\subsection{Size-Aware \SIEVE{}}
\label{sec:sieve-variants}

We study two causal size-aware \SIEVE{} variants: Mean-Threshold and
Dual-Threshold. Both keep the insertion-ordered resident queue and persistent
eviction hand of canonical \SIEVE{}. The only replacement-state change is that
the visited bit becomes a small nonnegative counter \(c_x\). A larger counter
means that the object can survive more hand examinations before eviction. The
counter does not predict reuse; it records the retention credit assigned from
the object's observed size and the current running mean size.

Let \(D_t\) be the set of distinct objects whose sizes have been observed before
or at request \(t\). Let \(L_x\) be the size of object \(x\). The common online
threshold is the running mean
\begin{equation}
  T_t = \frac{1}{|D_t|}\sum_{x\in D_t} L_x .
\label{eq:running-mean-size}
\end{equation}
The mean is updated causally as sizes are observed.

Mean-Threshold \SIEVE{} uses the running mean as a binary split:
\begin{equation}
\tau^{\mathrm{mean}}_t(L)=
\begin{cases}
2, & L < T_t,\\
1, & L \ge T_t .
\end{cases}
\label{eq:mean-token}
\end{equation}
A below-mean object receives two units of retention credit after a hit, so it
can survive two later hand examinations. An object at or above the mean receives
one unit after a hit, matching the one-extra-chance behavior of canonical
\SIEVE{}.

\begin{algorithm}[t]
\caption{Mean-Threshold \SIEVE{}}
\label{alg:mean-threshold-sieve}
\vspace{2pt}\hrule\vspace{4pt}
\begin{tabbing}
\quad\=\quad\=\kill
\textbf{State:} resident queue; hand; counter \(c_x\in\{0,1,2\}\); running mean \(T_t\).\\[1pt]
\textbf{On request for \(x\):} update \(D_t\) and \(T_t\) using observed size \(L_x\).\\[1pt]
\textbf{Token rule:} \(\tau_t(L_x)\gets2\) if \(L_x<T_t\); otherwise \(\tau_t(L_x)\gets1\).\\[1pt]
\textbf{On hit to \(x\):} \(c_x\gets\tau_t(L_x)\).\\[1pt]
\textbf{On miss for admissible \(x\):} insert \(x\) at the newest end with \(c_x\gets\max\{\tau_t(L_x)-1,0\}\).\\[2pt]
\textbf{While the cache exceeds capacity:}\\
\> \(y\gets\) object currently at the hand\\
\> \textbf{if} \(c_y=0\) \textbf{then} evict \(y\)\\
\> \textbf{else} \(c_y\gets c_y-1\) and advance the hand
\end{tabbing}
\vspace{2pt}\hrule
\end{algorithm}

Dual-Threshold \SIEVE{} uses the same mean but separates the right tail from the
middle of the distribution:
\begin{equation}
\tau^{\mathrm{dual}}_t(L)=
\begin{cases}
2, & L < T_t,\\
1, & T_t \le L < 2T_t,\\
0, & L \ge 2T_t .
\end{cases}
\label{eq:dual-token}
\end{equation}
Thus small objects receive extra retention credit, medium objects receive the
canonical one-hit protection, and very large objects receive no hit-based
retention credit after weak evidence. A large object can still be cached, but it
must compete without the extra hand pass that would otherwise let it occupy many
bytes after only one hit.

\begin{algorithm}[t]
\caption{Dual-Threshold \SIEVE{}}
\label{alg:dual-threshold-sieve}
\vspace{2pt}\hrule\vspace{4pt}
\begin{tabbing}
\quad\=\quad\=\kill
\textbf{State:} resident queue; hand; counter \(c_x\in\{0,1,2\}\); running mean \(T_t\).\\[1pt]
\textbf{On request for \(x\):} update \(D_t\) and \(T_t\) using observed size \(L_x\).\\[1pt]
\textbf{Token rule:}\\
\> \textbf{if} \(L_x<T_t\), set \(\tau_t(L_x)\gets2\)\\
\> \textbf{else if} \(L_x<2T_t\), set \(\tau_t(L_x)\gets1\)\\
\> \textbf{else} set \(\tau_t(L_x)\gets0\)\\[1pt]
\textbf{On hit to \(x\):} \(c_x\gets\tau_t(L_x)\).\\[1pt]
\textbf{On miss for admissible \(x\):} insert \(x\) at the newest end with \(c_x\gets\max\{\tau_t(L_x)-1,0\}\).\\[2pt]
\textbf{While the cache exceeds capacity:}\\
\> \(y\gets\) object currently at the hand\\
\> \textbf{if} \(c_y=0\) \textbf{then} evict \(y\)\\
\> \textbf{else} \(c_y\gets c_y-1\) and advance the hand
\end{tabbing}
\vspace{2pt}\hrule
\end{algorithm}

The insertion rule gives an object one fewer token than it would receive on a
hit. This preserves the canonical \SIEVE{} distinction between admission and
observed reuse: a newly admitted object is not fully protected until it is
requested again while resident. Mean-Threshold therefore protects below-mean
objects after either insertion or a later hit, while at-or-above-mean objects
need a hit before they survive the hand. Dual-Threshold adds a second correction
for the far tail: objects at least twice the running mean enter with zero
tokens and remain at zero even after a hit.

The direction of the token assignment follows the paper's primary objective,
object hit rate under a byte capacity. The rule does not assume that small
objects are intrinsically more valuable. It keeps large objects with weak reuse
evidence from occupying many bytes when those
bytes could instead serve many smaller requests. Compared with canonical
\SIEVE{}, the variants add the running mean and one extra counter bit per
resident object.

\subsection{Established Size-Aware Policies}
\label{sec:lru-components}

We also compare against three established size-aware policies:
LRU-MIN~\cite{WilliamsLRUMIN96}, AdaptSize~\cite{BergerAdaptSize17}, and
LHD~\cite{BeckmannLHD18}. We chose them because each adds object size at a
different point in cache management. LRU-MIN changes victim selection: it evicts
the least recently used object among those at least as large as the incoming
object. AdaptSize changes admission: it admits a missed object of size $s$ with
probability $e^{-s/c}$ and tunes $c$ online. LHD ranks resident objects by their
estimated hits per unit of cache space and time and evicts the lowest. We use these
policies without modification and report LRU-MIN as LRU\textsuperscript{!}.
Appendix~\ref{app:established} describes them in detail.

\subsection{Segmented and Blended Cache Policies}
\label{sec:blended-construction}

Many replacement policies separate newly admitted objects, in a \emph{cold}
region, from objects that have shown stronger evidence of reuse, in a
\emph{hot} region, and several keep \emph{ghost} histories of recently evicted
objects: a ghost hit is still a miss, but it shows that the object returned after
eviction. We build on three such policies, 2Q~\cite{JohnsonShasha94},
ARC~\cite{MegiddoModha03}, and S3-FIFO~\cite{YZQ23}. \SIEVE{} can also manage
one resident region inside such a policy~\cite{ZYY24}; we call the results
\emph{blended} policies, namely 2Q-SIEVE, ARC-SIEVE, and S3-FIFO-SIEVE. In each,
the parent policy keeps its admission path, ghost histories, partitioning rule,
and promotion semantics, and only its protected region is managed by \SIEVE{}.
How to divide an LRU cache among request classes has also been analyzed
directly~\cite{QuanMultiFlowLRU19}. Appendix~\ref{app:segmented} describes the
parent policies and the blended construction, and Table~\ref{tab:policy-roles}
there summarizes their structure.

\subsection{New Segmented Size-Aware Policies}
\label{sec:segmented-size-aware}

The segmented policies we evaluate are new. They are not the original 2Q, ARC,
and S3-FIFO, nor the blended policies of Section~\ref{sec:blended-construction}
used verbatim: we modify each of them to be size-aware, as described below.

The overall evaluation uses a compact notation: a superscript
\textsuperscript{!} marks the size-aware policy evaluated under a byte-capacity
budget. FIFO is ordinary FIFO: we do not change it, and it is included for
comparison without the superscript. LRU\textsuperscript{!} denotes LRU-MIN, and
\SIEVE{}\textsuperscript{!} denotes the Dual-Threshold \SIEVE{} rule from
Section~\ref{sec:sieve-variants}. For the segmented policies below, the
superscript denotes the corresponding byte-aware construction used in the replay
experiments.

2Q\textsuperscript{!} keeps 2Q's separation between a cold FIFO region and a
protected recency region. Requests that return through the ghost path indicate
that the object deserves stronger protection, while the resident replacement
step is made byte-aware so that large objects must justify their capacity cost.
2Q-SIEVE\textsuperscript{!} uses the same 2Q admission and ghost-hit structure,
but its protected region is governed by the size-aware \SIEVE{} rule rather than
an LRU ordering.

ARC\textsuperscript{!} uses the same structure as ARC: one resident list for
objects seen once, one for objects that have been requested again, and a ghost
list for each that remembers recently evicted objects. When a request hits in a
ghost list, the object is placed directly in the protected list. Eviction within
both resident lists takes object size into account, so a large object is not kept
at the expense of many small ones. ARC-SIEVE\textsuperscript{!} keeps this
structure and changes only the protected list, which is managed by the size-aware
\SIEVE{} rule instead of an LRU ordering.

S3-FIFO\textsuperscript{!} retains the S3-FIFO structure: a small FIFO filter, a
larger main FIFO region, and a ghost queue that identifies objects returning
after eviction. Reuse counters decide whether an object leaving the main region
is reinserted, and the size-aware form modulates that protection by the object's
size relative to the running mean. S3-FIFO-SIEVE\textsuperscript{!} keeps the
same admission filter and ghost behavior, but manages the protected region with
size-aware \SIEVE{}.

All six segmented policies are online. Their decisions use the current request,
resident metadata, ghost metadata maintained by the parent policy, and running
size statistics; they do not use future references or an offline trace pass.

\section{Experimental Methodology}
\label{sec:methodology}

\subsection{Production Traces}
\label{sec:production-traces}

\begin{figure*}[t]
\centering
\begin{minipage}{0.40\textwidth}
\centering
\includegraphics[width=\linewidth]{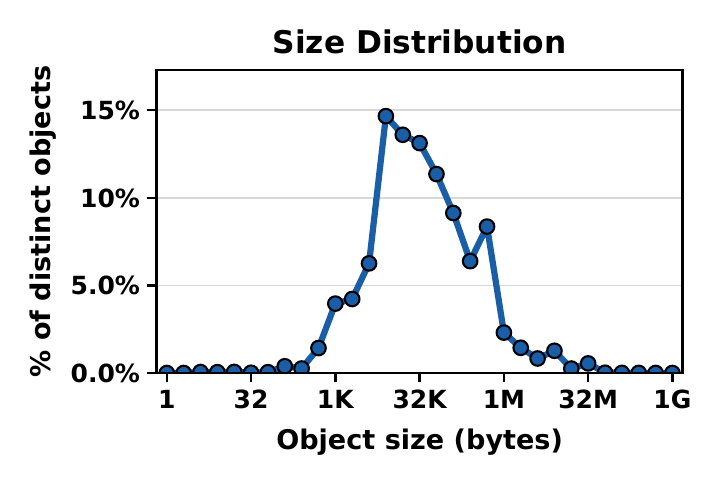}
{\small (a) META TRACE 1: size distribution}
\end{minipage}\hfill
\begin{minipage}{0.40\textwidth}
\centering
\includegraphics[width=\linewidth]{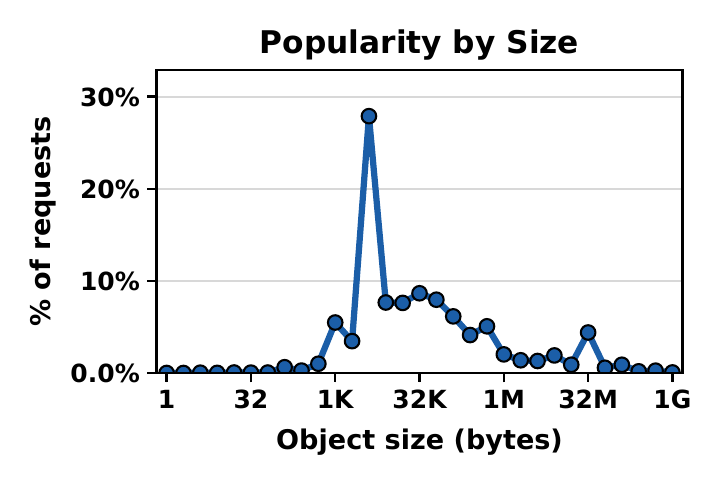}
{\small (b) META TRACE 1: popularity by size}
\end{minipage}
\caption{Size and request-popularity structure in META TRACE 1. Panel (a) shows
the distribution of distinct objects across object-size ranges, while panel (b)
shows how much request popularity is contributed by objects in those same size
ranges. Roughly speaking, size-aware policies are important when highly popular
objects whose requests are correlated in time vary substantially in size and
together exceed the available cache memory. A size-aware policy is therefore
most useful when the request mass is not confined to one narrow size range: then
objects with comparable reuse evidence can impose very different byte costs on
the cache. META TRACE 1 exhibits a broad size tail with substantial request mass
outside the smallest objects, creating strong pressure for size-aware retention.}
\Description{Two panels showing META TRACE 1 object-size distribution and
popularity by object size.}
\label{fig:size-dist}
\end{figure*}

We evaluate four production traces: three META production traces and one WIKI
TRACE~\cite{ProductionTraces}. The META traces are stored in the 
\texttt{oracleGeneral} format used by
CacheLib and libCacheSim~\cite{BergCacheLib20}; we refer to them as META TRACE
1, 2, or 3. The fourth production trace is WIKI TRACE.
Table~\ref{tab:workload} summarizes the four workloads.
Figure~\ref{fig:size-dist} shows both the object-size distribution and the
request popularity contributed by each size range for META TRACE 1;
Appendix~\ref{app:trace-characterization} gives the corresponding figures for
WIKI TRACE, META TRACE 2 and 3.

The production traces are useful because they expose different degrees of size
skew. For object $k$, let $L_k$ denote its size in bytes, $n_k$ the number of
requests for it, and $Q$ the total number of requests in the trace. We
write the empirical request frequency as
\begin{equation}
    q_k = \frac{n_k}{Q} .
    \label{eq:empirical-popularity}
\end{equation}
\textit{Size-aware replacement is most likely to help when frequently requested objects
have substantially different sizes, when the aggregate size of those competing
objects is larger than the cache budget $B$, and when their requests overlap in
time.} Equivalently, for a popularity cutoff $\pi$, the difficult regime is one
in which
$    \sum_{k:q_k \ge \pi} L_k > B $
while the corresponding objects $k$ are actively requested over similar periods of the trace.
Obviously, such a condition cannot be used by a feasible caching policy.

The four production traces span this regime to varying degrees. The three META traces
have multi-terabyte object footprints and heavy size tails: the largest $1\%$ of
distinct objects account for $53.2\%$--$68.9\%$ of the footprint. WIKI TRACE is
larger in request count but has a smaller footprint and a lighter tail, with the
largest $1\%$ of objects accounting for $42.4\%$ of total object size. These
contrasts let us test whether size-aware admission and protection are most
valuable when a small number of large objects can displace many smaller reusable
objects.

\begin{table*}[t]
\centering
\caption{Workload summary for the four production traces. Request and object
counts are in millions. Total object size is the sum over one copy of each
distinct object. The last column is the share of total object size held by the
largest $1\%$ of distinct objects.}
\label{tab:workload}
\small
\resizebox{\textwidth}{!}{%
\begin{tabular}{lrrrrrrr}
\toprule
Trace & Requests & Objects & Total object size & Median size & p99 size & Max size & Top $1\%$ share \\
\midrule
META TRACE 1 & 45.4\,M & 12.2\,M & 9.90\,TB & 42.5\,kB & 11.5\,MB & 2.11\,GB & 53.7\% \\
META TRACE 2 & 95.8\,M & 35.0\,M & 20.0\,TB & 33.8\,kB & 10.0\,MB & 2.13\,GB & 68.9\% \\
META TRACE 3 & 87.7\,M & 29.0\,M & 16.3\,TB & 34.9\,kB & 10.5\,MB & 2.14\,GB & 53.2\% \\
WIKI TRACE & 2803.5\,M & 37.2\,M & 5.80\,TB & 14.0\,kB & 3.42\,MB & 1.28\,GB & 42.4\% \\
\bottomrule
\end{tabular}}
\end{table*}

Figure~\ref{fig:size-dist} shows why the size distribution matters for this
study. The left column measures
how many distinct objects fall in each size range; the right column measures how
much of the request stream those objects generate. META TRACE 1 has both a long
object-size tail and meaningful popularity spread across multiple size ranges,
so a cache can face direct tradeoffs between retaining a large reusable object
and retaining many smaller reusable objects. The appendix gives the same
characterization for WIKI TRACE and for META TRACE 2 and 3. WIKI TRACE has a
lighter tail and a larger share of requests concentrated among smaller objects.
META TRACE 2 and 3 follow the same qualitative pattern as META TRACE 1,
with substantial size heterogeneity and nontrivial request mass outside
a single small-object bucket.

\subsection{Synthetic Traces}
\label{sec:synthetic-traces}

Production traces mix several effects: popularity skew, temporal locality,
object-size skew, and correlations between object size and request behavior. We
therefore also use synthetic traces to isolate one question: how does the value
of size-aware replacement change as object-size variability increases, while the
request process is held fixed?

Each synthetic trace has $N=10^6$ objects and $Q=10^7$ requests. Requests are
sampled independently from a Zipf distribution over object rank,
\begin{equation}
    p_k = \frac{k^{-\alpha}}{\sum_{j=1}^{N} j^{-\alpha}}, \qquad
    k \in \{1,\ldots,N\},
    \label{eq:zipf}
\end{equation}
with $\alpha=0.6$.
This relatively small Zipf exponent keeps a
large set of objects warm rather than concentrating almost all requests on a
small head. 
Object sizes are then sampled independently of popularity, so
that the synthetic workload does not build in a size--popularity correlation
that favors size-aware policies; any improvement must come
from handling size variability under the same request stream.

Let $\mu_L$ be the mean object size and $\sigma_L$ its (empirical) standard deviation,
\begin{equation}
    \mu_L = \frac{1}{N}\sum_{k=1}^{N} L_k, \qquad
    \sigma_L^2 = \frac{1}{N}\sum_{k=1}^{N} (L_k-\mu_L)^2 .
    \label{eq:size-moments}
\end{equation}
We report variability using the coefficient of variation,
\begin{equation}
    \mathrm{CV}(L) = \left. \sigma_L\middle/ \mu_L \right.,
    \label{eq:cv}
\end{equation}
We use four synthetic traces, which differ only in their object-size
distribution. The two uniform distributions, $U(1,100)$ and $U(1,500)$, draw
sizes uniformly from $1$ to $100$ and from $1$ to $500$ bytes. They are low-CV
controls: their mean sizes differ ($50.5$ and $250.4$ bytes), but their
coefficients of variation are nearly the same ($0.57$ and $0.58$). The two
lognormal distributions, written LN(CV=2) and LN(CV=4), keep the mean near
$50$ bytes ($50.0$ and $51.4$) and raise the coefficient of variation to $1.99$
and $3.87$. Lognormal samples are
bounded by rejection sampling over the allowed range rather than by clamping; in
particular, values outside the range are resampled instead of being accumulated
at an artificial maximum-size point. Table~\ref{tab:synthetic-workloads} in
Appendix~\ref{app:synthetic-traces} lists the mean, standard deviation, and
coefficient of variation of each trace.

If popular
objects are nearly equal in size, a size-aware policy has little extra
information beyond reuse. If popular objects have a broad size distribution and
cannot all fit in the cache, the eviction decision becomes a byte-allocation
problem: protecting one very large object can evict many smaller objects with
similar request rates. The synthetic traces make this transition explicit by
keeping the popularity process fixed while increasing the coefficient of
variation of the size distribution.

\subsection{Cache Sizes and Metrics}

Cache sizes are fixed byte budgets. We use $0.01\%$,
$0.1\%$, $1\%$, $5\%$, $10\%$, $20\%$, and $30\%$ of each trace's total object
size. Object sizes are
read directly from the trace records. Records whose size is unknown are skipped by every policy. The
oracle next-access field of the production records is never passed to any
policy, so no evaluated online policy can use future reuse.

The largest cache sizes, $20\%$ and $30\%$ of the total object size, are larger
than those usually studied. They may be realistic for edge or near-edge caches that serve
real-time applications, such as virtual or augmented reality, where the data must
be close to the user when it is requested. Also, $20\%$--$30\%$ 
still represents
significant savings considering the cost of edge  versus remote
cloud storage.

We evaluate policies in terms of object hit rate 
and average runtime per request (cache policy
overhead), noting the latter may be pipelined with
object I/O overhead.
Byte hit rate performance is given in the Appendix.


\section{Size-Aware \SIEVE{}}
\label{sec:sieve-results}

We first isolate the effect of the size signal in standalone \SIEVE{}. The
comparison includes canonical \SIEVE{}, MT-SIEVE, and DT-SIEVE. MT-SIEVE is
Mean-Threshold \SIEVE{}, and DT-SIEVE is Dual-Threshold \SIEVE{}. The same
color mapping is used in every figure of the paper: canonical \SIEVE{} is gray,
MT-SIEVE is blue, and DT-SIEVE is red; the same red marks
\SIEVE{}\textsuperscript{!} in Section~\ref{sec:overall-results}, since it is the
same policy.

\subsection{Production Traces}
\label{sec:sieve-production}

\begin{figure*}[t]
\centering
\includegraphics[width=0.40\textwidth]{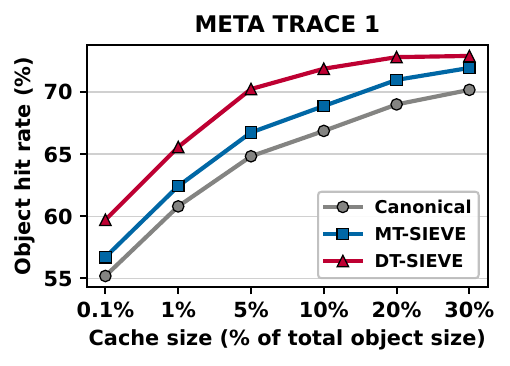}\hfill
\includegraphics[width=0.40\textwidth]{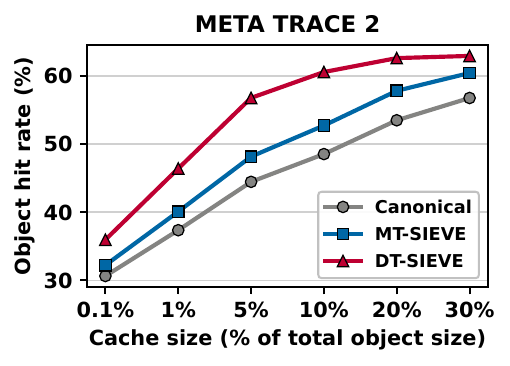}\\[0.6em]
\includegraphics[width=0.40\textwidth]{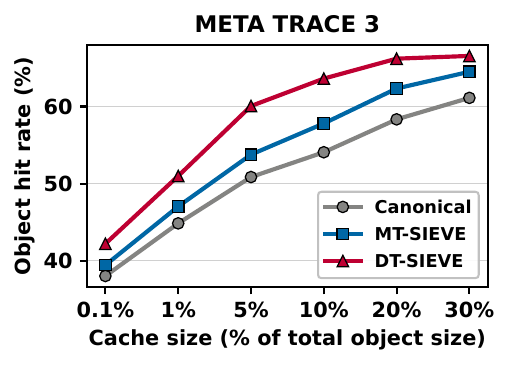}\hfill
\includegraphics[width=0.40\textwidth]{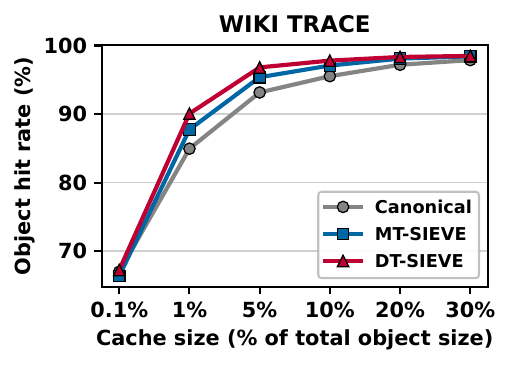}
\caption{Object hit rate for canonical \SIEVE{}, MT-SIEVE, and DT-SIEVE on the
three META traces and WIKI TRACE. MT-SIEVE means Mean-Threshold \SIEVE{}, which
adds extra retention credit to objects smaller than the running mean size.
DT-SIEVE means Dual-Threshold \SIEVE{}, which also removes hit-based protection
from objects at least twice the running mean size. The y-axis is the fraction of
requests served directly from cache; higher values imply fewer backend requests
for users at the same byte capacity.}
\Description{Four line charts, one per production trace, showing object hit rate versus cache size for canonical SIEVE, MT-SIEVE, and DT-SIEVE.}
\label{fig:sieve-real-ohr}
\end{figure*}

Table~\ref{tab:sieve-main} reports object hit rate at every measured cache
size for the three META traces. It also reports the relative improvement of
MT-SIEVE and DT-SIEVE over canonical \SIEVE{} at the same cache size; these
gains are percentages. Figure~\ref{fig:sieve-real-ohr} shows the full
production sweep, including WIKI TRACE. DT-SIEVE is the strongest policy in
every row of the table. At a $10\%$ cache in the full production sweep, it
increases object hit rate over canonical \SIEVE{} by $7.5\%$ on META TRACE 1,
$24.8\%$ on META TRACE 2, $17.7\%$ on META TRACE 3, and $2.4\%$ on WIKI TRACE.

\begin{table*}[t]
\centering
\caption{META-trace object hit rate for canonical \SIEVE{} and the two
size-aware variants. MT-SIEVE denotes Mean-Threshold \SIEVE{}, and DT-SIEVE
denotes Dual-Threshold \SIEVE{}. OHR columns are object hit rates in percent;
higher OHR means more user requests are served from cache instead of going to the
backend. Gain columns report relative percentage improvement over canonical
\SIEVE{} at the same trace and cache size. Best OHR in each row is bold.}
\label{tab:sieve-main}
\scriptsize
\setlength{\tabcolsep}{3pt}
\resizebox{\textwidth}{!}{%
\begin{tabular}{llrrrrr}
\toprule
Trace & Cache & Canonical OHR & MT-SIEVE OHR & MT-SIEVE gain & DT-SIEVE OHR & DT-SIEVE gain \\
\midrule
META TRACE 1 & 0.01\% & 49.93 & 51.11 & +2.3\% & \cellcolor{gray!18}\textbf{53.05} & +6.2\% \\
 & 0.1\% & 55.19 & 56.70 & +2.7\% & \cellcolor{gray!18}\textbf{59.71} & +8.2\% \\
 & 1\% & 60.79 & 62.42 & +2.7\% & \cellcolor{gray!18}\textbf{65.57} & +7.9\% \\
 & 5\% & 64.82 & 66.74 & +3.0\% & \cellcolor{gray!18}\textbf{70.24} & +8.3\% \\
 & 10\% & 66.86 & 68.87 & +3.0\% & \cellcolor{gray!18}\textbf{71.87} & +7.5\% \\
 & 20\% & 69.00 & 70.97 & +2.9\% & \cellcolor{gray!18}\textbf{72.80} & +5.5\% \\
 & 30\% & 70.17 & 71.93 & +2.5\% & \cellcolor{gray!18}\textbf{72.90} & +3.9\% \\
\midrule
META TRACE 2 & 0.01\% & 24.01 & 25.22 & +5.0\% & \cellcolor{gray!18}\textbf{26.55} & +10.6\% \\
 & 0.1\% & 30.61 & 32.20 & +5.2\% & \cellcolor{gray!18}\textbf{35.94} & +17.4\% \\
 & 1\% & 37.34 & 40.07 & +7.3\% & \cellcolor{gray!18}\textbf{46.40} & +24.3\% \\
 & 5\% & 44.44 & 48.15 & +8.3\% & \cellcolor{gray!18}\textbf{56.77} & +27.7\% \\
 & 10\% & 48.52 & 52.71 & +8.6\% & \cellcolor{gray!18}\textbf{60.57} & +24.8\% \\
 & 20\% & 53.48 & 57.81 & +8.1\% & \cellcolor{gray!18}\textbf{62.62} & +17.1\% \\
 & 30\% & 56.76 & 60.43 & +6.5\% & \cellcolor{gray!18}\textbf{62.94} & +10.9\% \\
\midrule
META TRACE 3 & 0.01\% & 30.29 & 31.31 & +3.4\% & \cellcolor{gray!18}\textbf{32.75} & +8.1\% \\
 & 0.1\% & 38.02 & 39.43 & +3.7\% & \cellcolor{gray!18}\textbf{42.20} & +11.0\% \\
 & 1\% & 44.84 & 47.03 & +4.9\% & \cellcolor{gray!18}\textbf{50.99} & +13.7\% \\
 & 5\% & 50.85 & 53.76 & +5.7\% & \cellcolor{gray!18}\textbf{60.06} & +18.1\% \\
 & 10\% & 54.07 & 57.80 & +6.9\% & \cellcolor{gray!18}\textbf{63.62} & +17.7\% \\
 & 20\% & 58.33 & 62.34 & +6.9\% & \cellcolor{gray!18}\textbf{66.21} & +13.5\% \\
 & 30\% & 61.12 & 64.51 & +5.5\% & \cellcolor{gray!18}\textbf{66.56} & +8.9\% \\
\bottomrule
\end{tabular}
}
\end{table*}

The result is consistent with the size distributions in
Table~\ref{tab:workload}. The META traces contain a small number of very large
objects that can consume a large share of the byte budget. Mean-Threshold uses
the running mean to give smaller objects an additional hand pass, which already
improves over canonical \SIEVE{}. Dual-Threshold keeps that protection for
below-mean objects and also prevents objects at least twice the mean size from
receiving hit-based protection after weak evidence. The dual rule is therefore
especially effective on META TRACE 2 and 3, where large objects exert
stronger pressure on the resident byte budget.

WIKI TRACE also benefits, but the gap is smaller. Its object-size distribution
is less extreme than the META traces, so there are fewer large objects whose
removal can free enough space for many smaller reusable objects. The direction,
however, remains the same: the size-aware variants increase request hits.

\subsection{Runtime and Byte Hit Rate}
\label{sec:sieve-runtime}

\begin{figure}[t]
\centering
\includegraphics[width=0.60\linewidth]{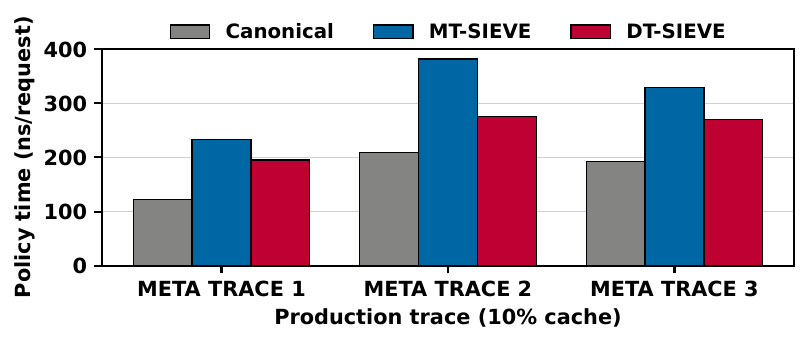}
\caption{Wall-clock replacement-policy overhead for canonical \SIEVE{},
MT-SIEVE, and DT-SIEVE on the three META traces at the $10\%$ cache size.
MT-SIEVE is Mean-Threshold \SIEVE{} and DT-SIEVE is Dual-Threshold \SIEVE{}.
Values are simulator policy time in nanoseconds per request; lower bars mean
less CPU and metadata overhead for cache management. This timing excludes
backend service latency and object transfer time, which can be much larger for
variable-length objects.}
\Description{Grouped bar chart of policy time per request for canonical SIEVE, MT-SIEVE, and DT-SIEVE on the three META traces at the 10 percent cache size.}
\label{fig:sieve-wall-clock}
\end{figure}

Figure~\ref{fig:sieve-wall-clock} reports whole-trace policy time for the three
META traces at the $10\%$ cache size. We use this replay time as a measure of
replacement-policy overhead: it is the wall-clock time required by the simulator per
request to apply the policy logic, not an end-to-end storage or network
latency. Lower values mean less controller CPU and metadata work. The
size-aware policies add one running mean update and a small token calculation to
the \SIEVE{} path, so they are slower than canonical \SIEVE{} in this replay
engine; nevertheless, the absolute times remain in the hundreds of nanoseconds
per request. The full META timing table, including the $30\%$ cache size, is in
Appendix~\ref{app:sieve}. In a deployed cache, this eviction overhead can be
pipelined with the memory or disk I/O that moves object payloads. That I/O can
be substantial for very large objects, so part of the extra policy time may be
hidden rather than added to request latency.

Byte hit rate measures the fraction of requested bytes served from cache. It reflects the efficiency of cache capacity usage.
It is difficult to discriminate different cache-eviction policies according
to this criterion, see Figure~\ref{fig:app-sieve-bhr} in
Appendix~\ref{app:sieve}. Prior work such as AdaptSize
also reports no noticeable difference in disk-cache byte hit rate
~\cite{BergerAdaptSize17}.

\subsection{Synthetic Traces}
\label{sec:sieve-synthetic}

\begin{figure*}[t]
\centering
\includegraphics[width=0.40\textwidth]{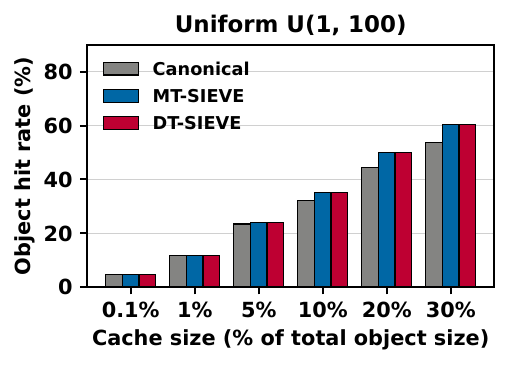}\hfill
\includegraphics[width=0.40\textwidth]{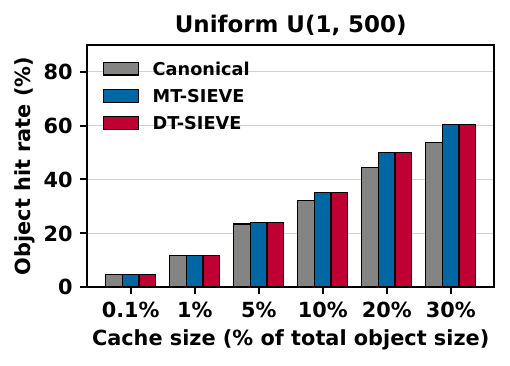}\\[0.6em]
\includegraphics[width=0.40\textwidth]{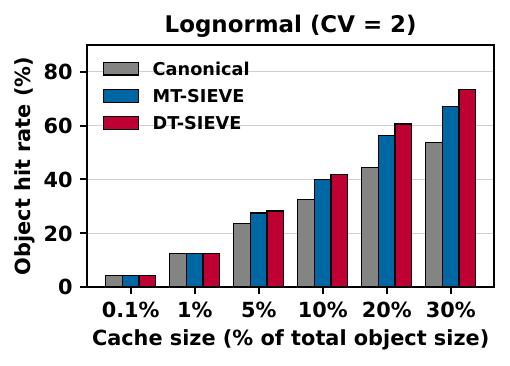}\hfill
\includegraphics[width=0.40\textwidth]{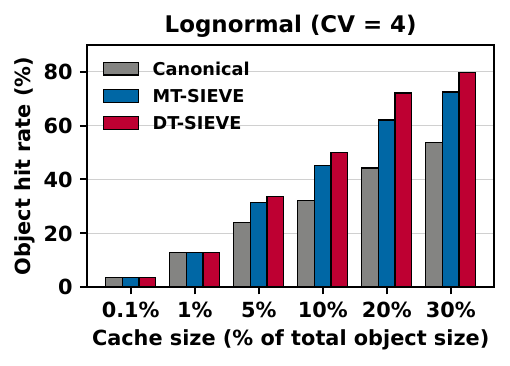}
\caption{Object hit rate of canonical \SIEVE{}, MT-SIEVE (Mean-Threshold \SIEVE{}),
and DT-SIEVE (Dual-Threshold \SIEVE{}) on the four synthetic traces, at cache sizes of $0.1\%$--$30\%$ of the total
object size. Lognormal (CV = 2) and Lognormal (CV = 4) are written LN(CV=2) and
LN(CV=4) in the text, where CV is the coefficient of variation of object size.
All panels use the same y-axis range.}
\Description{Four bar charts, one per synthetic trace, comparing the object hit
rate of canonical SIEVE, MT-SIEVE, and DT-SIEVE across six cache sizes.}
\label{fig:sieve-synthetic}
\end{figure*}

We use synthetic traces to separate the effect of the size distribution from the
production traces' other workload properties. Figure~\ref{fig:sieve-synthetic}
shows the four synthetic settings used in the main paper: \(U(1,100)\),
\(U(1,500)\), LN(CV=2), and LN(CV=4), all with Zipf parameter \(\alpha=0.6\),
see \eqref{eq:zipf}.

The two uniform traces are the control case. Moving from \(U(1,100)\) to
\(U(1,500)\) widens the absolute size range, but it does not make objects more
different relative to a typical object: both traces have CV about \(0.58\).
Consequently, the running mean does not isolate a much heavier tail, and the hit
rates barely move. The lognormal traces create the opposite situation. Their
mean object size stays near the same scale, but their relative dispersion is much
larger. At the $10\%$ cache size, increasing CV from about \(2\) to about \(4\)
raises DT-SIEVE's relative gain over canonical \SIEVE{} from about \(29\%\) to
about \(55\%\). A wider size range alone therefore does not make size awareness
pay off; what matters is a heavy relative tail, in which a few large objects can
displace many smaller reused objects.

At the smallest cache sizes the three variants perform almost equally: at $0.1\%$ and
$1\%$ of the total object size their object hit rates are within $0.4\%$ of each
other on all four traces. The cache then holds only about $1{,}000$ or $10{,}000$
of the $10^6$ objects. As Figure~\ref{fig:sieve-synthetic} shows, the
size-aware variants separate from canonical \SIEVE{} from the $5\%$ cache size
onward. Table~\ref{tab:app-sieve-synthetic} in Appendix~\ref{app:sieve} lists
the values and the relative gains at every cache size; the largest gain is
$62.9\%$, for DT-SIEVE on LN(CV=4) at the $20\%$ cache size.

\subsection{Why the Size-Aware Rule Helps}

Canonical \SIEVE{} treats the first hit to a small object and the first hit to a
large object identically. Under a byte budget this can be costly: a
large object with weak reuse evidence can displace many smaller objects that
would each have served later requests. Mean-Threshold changes this by
giving below-mean objects one additional hand pass. Dual-Threshold adds a second
correction for the far tail by withholding hit-based protection from objects at
least twice the running mean size.

Both rules are local. The policy observes object sizes as requests arrive and changes
only the amount of retention credit assigned when objects are inserted, hit, and
examined by the hand.

\section{Overall Results}
\label{sec:overall-results}

We now compare our proposed size-aware policies with existing size-aware
policies across the production and synthetic workloads. Our proposed policies are
\SIEVE{}\textsuperscript{!} (Dual-Threshold \SIEVE{}) and the size-aware
segmented versions of 2Q, 2Q-SIEVE, ARC, ARC-SIEVE, S3-FIFO, and S3-FIFO-SIEVE,
written with the superscript of Section~\ref{sec:segmented-size-aware}. The
existing size-aware policies are LRU\textsuperscript{!} (LRU-MIN), LHD, and
AdaptSize (Section~\ref{sec:lru-components}). FIFO is included as a
size-oblivious reference. 

\subsection{Production Traces}
\label{sec:overall-production}

\begin{figure*}[t]
\centering
\includegraphics[width=0.40\textwidth]{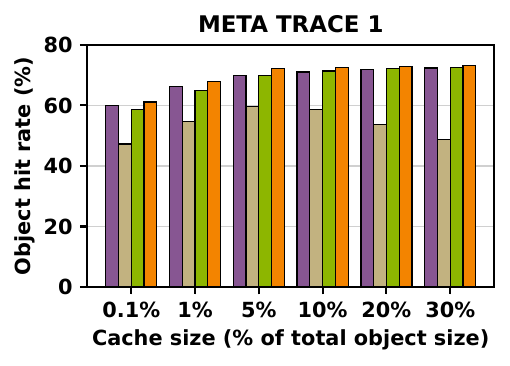}\hfill
\includegraphics[width=0.40\textwidth]{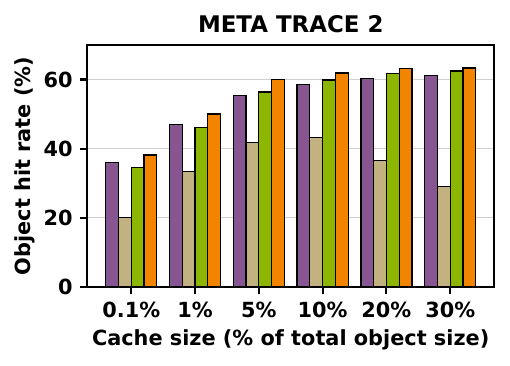}\\[0.3em]
\includegraphics[width=\textwidth]{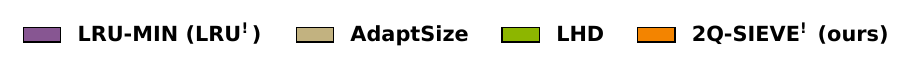}
\caption{Object hit rate of 2Q-SIEVE\textsuperscript{!}
(Section~\ref{sec:segmented-size-aware})
and the existing size-aware policies LRU-MIN (LRU\textsuperscript{!}), AdaptSize,
and LHD on META TRACE 1 and 2, at cache sizes from $0.1\%$ to $30\%$ of
the total object size. Our size-aware policies perform very well on these traces,
and among them 2Q-SIEVE\textsuperscript{!} has the highest object hit rate at almost
every cache size. It is higher than all three existing size-aware policies at every
cache size shown.}
\Description{Two grouped bar charts, for META TRACE 1 and 2, comparing the object hit rate of LRU-MIN, AdaptSize, LHD, and 2Q-SIEVE across six cache sizes.}
\label{fig:best-vs-existing}
\end{figure*}

Figure~\ref{fig:best-vs-existing} compares 2Q-SIEVE\textsuperscript{!},
our strongest policy on the production traces, with the
three existing size-aware policies, LRU-MIN, AdaptSize, and LHD, at every cache
size from $0.1\%$ to $30\%$ on META TRACE 1 and META TRACE 2.
2Q-SIEVE\textsuperscript{!} is a new policy developed in this paper
(Section~\ref{sec:segmented-size-aware}). It keeps the admission path and ghost
history of 2Q and manages the protected region with our size-aware
Dual-Threshold \SIEVE{} rule. On META TRACE 1 and 2,
2Q-SIEVE\textsuperscript{!} has a higher object hit rate than all three existing
size-aware policies at every one of these cache sizes, and META TRACE 3 shows the
same pattern (Table~\ref{tab:section5-production-ohr}). Its lead over the best
existing policy (LRU-MIN or LHD, depending on the trace and cache size) is largest
when the cache is small, $1.8\%$--$6.7\%$ at cache sizes of $0.1\%$ to $5\%$, and
it remains $0.9\%$--$1.3\%$ at $30\%$. Against AdaptSize the gap is
$21\%$--$118\%$.

\begin{figure*}[t]
\centering
\includegraphics[width=\textwidth]{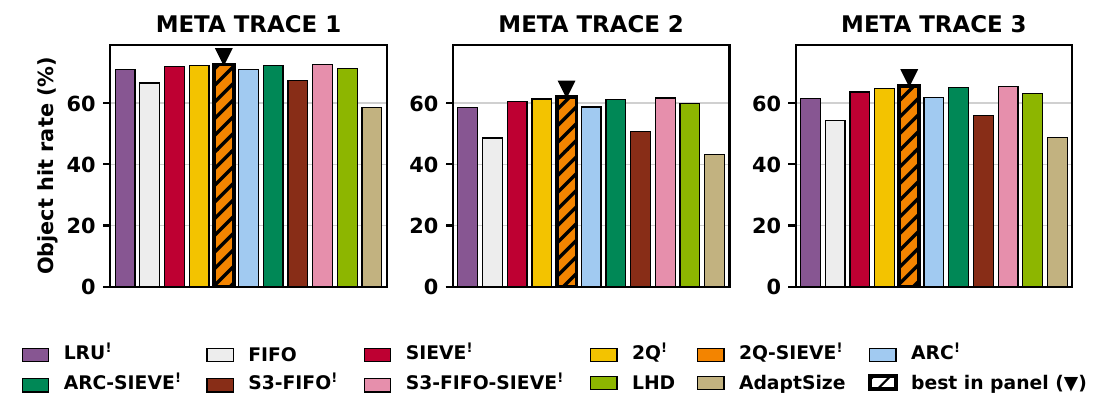}
\caption{META-trace object hit rate at the $10\%$ cache size for our proposed
policies and the existing size-aware policies (LRU\textsuperscript{!}, LHD, and AdaptSize). Object hit rate is the fraction of user
requests served from cache; higher bars mean fewer backend requests at the same
byte capacity. The hatched bar marks the best measured policy within each trace.
Each policy has the same color in every figure of the paper.}
\Description{Three bar charts, one per META trace, showing the object hit rate of our proposed policies and the existing size-aware policies at the 10 percent cache size.}
\label{fig:section5-production-ohr}
\end{figure*}

Figure~\ref{fig:section5-production-ohr} shows object hit rate at $10\%$ cache
size on the three META traces. The color assigned to each policy is fixed across
all Section~\ref{sec:overall-results} figures; the hatched bar is the best
policy within that trace. Table~\ref{tab:section5-production-ohr} reports every
META cache size.

\begin{table*}[t]
\centering
\caption{Object hit rate on the META production traces. Each entry is an object hit rate in percent; higher values mean a larger fraction of user requests are served directly from cache rather than sent to the backend. The table compares our proposed policies with the existing size-aware policies (LRU\textsuperscript{!}, i.e., LRU-MIN, LHD, and AdaptSize); the best policy in each row is bold.}
\label{tab:section5-production-ohr}
\scriptsize
\setlength{\tabcolsep}{1.8pt}
\resizebox{\textwidth}{!}{%
\begin{tabular}{llrrrrrrrrrrr}
\toprule
Trace & Cache & LRU\textsuperscript{!} & AdaptSize & LHD & FIFO & SIEVE\textsuperscript{!} & 2Q\textsuperscript{!} & 2Q-SIEVE\textsuperscript{!} & ARC\textsuperscript{!} & ARC-SIEVE\textsuperscript{!} & S3-FIFO\textsuperscript{!} & S3-FIFO-SIEVE\textsuperscript{!} \\
\midrule
META TRACE 1 & 0.01\% & \cellcolor{gray!18}\textbf{53.9} & 41.4 & 52.6 & 49.8 & 53.0 & 53.6 & 53.5 & 53.8 & 53.8 & 49.9 & 53.5 \\
 & 0.1\% & 60.1 & 47.2 & 58.7 & 55.2 & 59.7 & 61.0 & \cellcolor{gray!18}\textbf{61.2} & 59.9 & 60.8 & 55.9 & 61.1 \\
 & 1\% & 66.2 & 54.7 & 64.9 & 60.7 & 65.6 & 67.9 & \cellcolor{gray!18}\textbf{67.9} & 66.0 & 67.6 & 61.3 & 67.7 \\
 & 5\% & 69.9 & 59.6 & 70.0 & 64.8 & 70.2 & 71.3 & \cellcolor{gray!18}\textbf{72.1} & 69.9 & 71.7 & 65.5 & 72.0 \\
 & 10\% & 71.0 & 58.6 & 71.4 & 66.6 & 71.9 & 72.3 & \cellcolor{gray!18}\textbf{72.6} & 71.1 & 72.3 & 67.5 & 72.5 \\
 & 20\% & 72.0 & 53.8 & 72.2 & 68.8 & 72.8 & 72.8 & \cellcolor{gray!18}\textbf{72.9} & 72.2 & 72.7 & 69.3 & 72.8 \\
 & 30\% & 72.4 & 48.8 & 72.5 & 70.0 & 72.9 & 73.0 & \cellcolor{gray!18}\textbf{73.1} & 72.8 & 72.8 & 70.4 & 72.9 \\
META TRACE 2 & 0.01\% & \cellcolor{gray!18}\textbf{27.7} & 13.1 & 26.8 & 24.2 & 26.5 & 27.4 & 27.0 & 27.5 & 27.3 & 23.5 & 26.6 \\
 & 0.1\% & 35.9 & 20.1 & 34.6 & 31.2 & 35.9 & \cellcolor{gray!18}\textbf{38.5} & 38.1 & 35.7 & 37.2 & 30.7 & 38.0 \\
 & 1\% & 46.9 & 33.4 & 46.1 & 37.8 & 46.4 & \cellcolor{gray!18}\textbf{50.6} & 50.1 & 46.7 & 49.7 & 38.1 & 49.9 \\
 & 5\% & 55.4 & 41.7 & 56.4 & 44.8 & 56.8 & 58.7 & \cellcolor{gray!18}\textbf{60.1} & 55.4 & 59.6 & 46.0 & 60.0 \\
 & 10\% & 58.6 & 43.2 & 59.8 & 48.6 & 60.6 & 61.3 & \cellcolor{gray!18}\textbf{61.9} & 58.7 & 61.3 & 50.6 & 61.7 \\
 & 20\% & 60.3 & 36.5 & 61.7 & 53.5 & 62.6 & 63.0 & \cellcolor{gray!18}\textbf{63.1} & 61.9 & 62.4 & 55.0 & 62.6 \\
 & 30\% & 61.2 & 29.1 & 62.5 & 56.4 & 62.9 & 63.2 & \cellcolor{gray!18}\textbf{63.3} & 62.8 & 62.8 & 57.2 & 63.0 \\
META TRACE 3 & 0.01\% & 33.0 & 18.9 & 32.6 & 30.7 & 32.7 & 32.9 & 33.0 & 32.9 & \cellcolor{gray!18}\textbf{33.5} & 30.0 & 32.8 \\
 & 0.1\% & 40.7 & 25.1 & 40.8 & 38.1 & 42.2 & 43.1 & \cellcolor{gray!18}\textbf{44.1} & 40.5 & 43.3 & 38.3 & 43.9 \\
 & 1\% & 49.7 & 40.5 & 50.8 & 44.7 & 51.0 & 53.6 & \cellcolor{gray!18}\textbf{54.4} & 49.5 & 53.9 & 45.5 & 54.2 \\
 & 5\% & 58.6 & 48.4 & 59.7 & 50.9 & 60.1 & 61.9 & \cellcolor{gray!18}\textbf{63.8} & 58.5 & 63.4 & 51.9 & 63.7 \\
 & 10\% & 61.5 & 48.8 & 63.2 & 54.2 & 63.6 & 64.7 & \cellcolor{gray!18}\textbf{65.6} & 61.9 & 65.1 & 55.9 & 65.5 \\
 & 20\% & 64.1 & 42.3 & 65.2 & 58.4 & 66.2 & 66.0 & \cellcolor{gray!18}\textbf{66.3} & 64.4 & 66.0 & 59.7 & 66.2 \\
 & 30\% & 65.1 & 33.6 & 65.9 & 60.8 & 66.6 & 66.7 & \cellcolor{gray!18}\textbf{66.8} & 66.2 & 66.4 & 61.5 & 66.6 \\
\bottomrule
\end{tabular}%
}
\end{table*}

2Q-SIEVE\textsuperscript{!} gives the highest object hit rate on the META traces
at most cache sizes. At the $10\%$ cache size of
Figure~\ref{fig:section5-production-ohr}, it reaches $72.6\%$, $61.9\%$, and
$65.6\%$ on META TRACE 1, 2, and 3. This is $1.6\%$--$3.9\%$ higher than LHD, the
strongest existing size-aware policy, $2.2\%$--$6.7\%$ higher than LRU-MIN, and
$24\%$--$43\%$ higher than AdaptSize. It is also $1.0\%$--$3.1\%$ higher than
standalone \SIEVE{}\textsuperscript{!}, so the segmented structure adds to the
size-aware rule rather than duplicating it.

2Q-SIEVE\textsuperscript{!} works well because it combines two complementary
filters. The 2Q path uses ghost history to identify objects that return after eviction,
while the size-aware fast path can admit small objects directly into the
protected region. Thus, protection is based on both observed reuse evidence and
the current object-size signal. The Dual-Threshold \SIEVE{} rule then decides how to allocate the protected
bytes among objects that have shown reuse: small reused objects receive extra
hand passes, and very large objects receive no hit-based protection. The first
filter removes objects that would never be requested again, and the second keeps
a few large objects from occupying bytes that many smaller reused objects could
use. Both matter on the META traces, where a small fraction of very large objects
holds most of the bytes.

Each existing size-aware policy has only one of these two filters. LRU-MIN uses
size only to choose eviction victims and relies on recency alone as reuse
evidence, so an object requested once still enters the recency order like any
other. AdaptSize uses size only at admission and admits large objects with small
probability even when they are requested repeatedly. LHD ranks objects by their
estimated hits per unit of space and time, but it admits every object and has no
ghost history to keep objects that are requested only once out of the cache.
2Q-SIEVE\textsuperscript{!} uses reuse evidence to decide which objects to
protect and object size to decide how strongly to protect them, which is why it
leads at every cache size from $0.1\%$ upward.

The other size-aware \SIEVE{}-based policies follow the same pattern. Replacing
the protected region of the parent policy with size-aware \SIEVE{} raises object
hit rate: at the $10\%$ cache size, ARC-SIEVE\textsuperscript{!} is
$1.7\%$--$5.1\%$ higher than ARC\textsuperscript{!}, and
S3-FIFO-SIEVE\textsuperscript{!} is $7.5\%$--$21.8\%$ higher than
S3-FIFO\textsuperscript{!} and within $0.3\%$ of 2Q-SIEVE\textsuperscript{!}. In
each case, the parent policy still supplies the admission and ghost evidence,
while size-aware \SIEVE{} decides how that evidence is converted into protection
under a byte budget.

At the smallest cache size, $0.01\%$ of the total object size, LRU-MIN is slightly
ahead on META TRACE 1 and 2. This cache is only about $1$--$2$\,GB, smaller than the largest objects in
the trace, and holds only a few thousand objects of average size. A single large
object can then take most of the cache, so removing large objects as early as
possible matters most, which is what LRU-MIN does by evicting objects at least
as large as the incoming one first. Our policies still admit every object that
fits and only withhold extra protection from large objects, so a large object
stays until the hand reaches it. Once the cache holds more objects, from $0.1\%$
upward, reuse evidence matters more and our policies take the lead.

\subsection{Synthetic Traces}
\label{sec:overall-synthetic}

\begin{figure*}[b]
\centering
\includegraphics[width=0.40\textwidth]{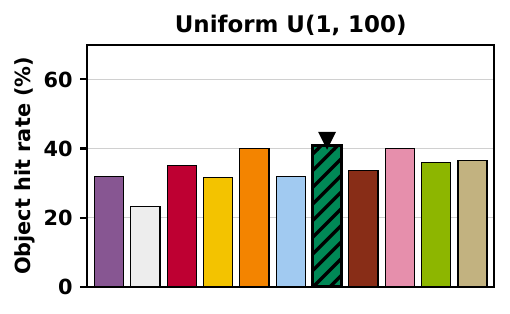}\hfill
\includegraphics[width=0.40\textwidth]{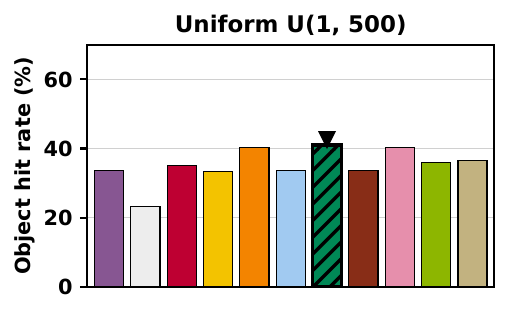}\\[0.6em]
\includegraphics[width=0.40\textwidth]{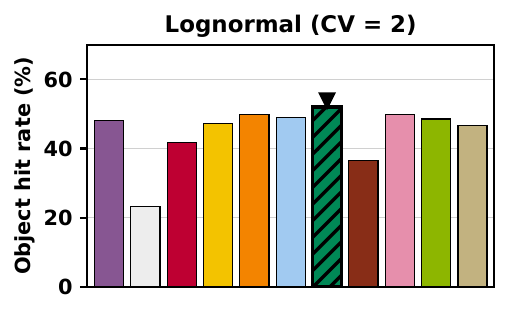}\hfill
\includegraphics[width=0.40\textwidth]{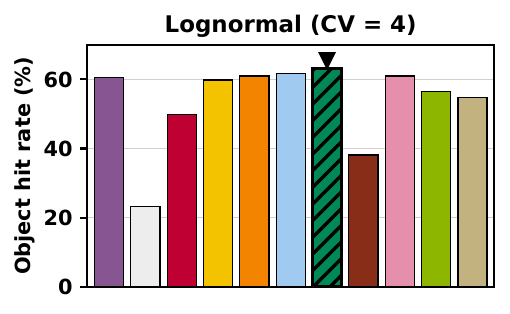}\\[0.3em]
\includegraphics[width=\textwidth]{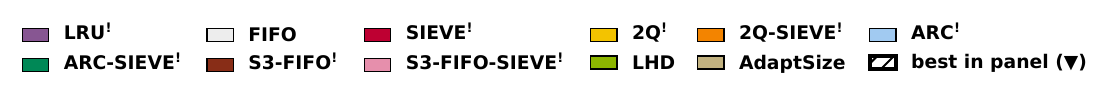}
\caption{Synthetic-trace object hit rate at the $10\%$ cache size and
Zipf parameter $\alpha=0.6$ for our proposed policies and the existing size-aware policies (LRU\textsuperscript{!}, LHD, and AdaptSize). CV denotes
the coefficient of variation, so the Lognormal (CV = 2) and Lognormal (CV = 4)
panels, written LN(CV=2) and LN(CV=4) in the text and tables, isolate relative
size dispersion. All four panels use the same y-axis range. Higher bars mean a larger fraction of requests are
served from cache. The hatched bar marks the best policy in each synthetic
trace, highlighting how the winner changes as the object-size distribution
becomes more variable.}
\Description{Four bar charts, one per synthetic trace, showing the object hit rate of our proposed policies and the existing size-aware policies at the 10 percent cache size.}
\label{fig:section5-synthetic-ohr}
\end{figure*}

Figure~\ref{fig:section5-synthetic-ohr} repeats the comparison on four synthetic
traces at Zipf exponent $\alpha=0.6$: $U(1,100)$, $U(1,500)$, LN(CV=2), and
LN(CV=4). Table~\ref{tab:section5-synthetic-ohr}
reports cache sizes $1\%$, $5\%$, $10\%$, $20\%$, and $30\%$.

\begin{table*}[t]
\centering
\caption{Object hit rate on four measured synthetic traces at $\alpha=0.6$ for cache sizes $1\%$, $5\%$, $10\%$, $20\%$, and $30\%$. At $0.1\%$ the cache holds only about $1{,}000$ of the $10^6$ objects and the object hit rate of every policy is at most $8\%$ (Table~\ref{tab:app-section5-synthetic-ohr-full}). The table compares our proposed policies with the existing size-aware policies (LRU\textsuperscript{!}, i.e., LRU-MIN, LHD, and AdaptSize); the best policy in each row is bold.}
\label{tab:section5-synthetic-ohr}
\scriptsize
\setlength{\tabcolsep}{1.8pt}
\resizebox{\textwidth}{!}{%
\begin{tabular}{llrrrrrrrrrrr}
\toprule
Trace & Cache & LRU\textsuperscript{!} & AdaptSize & LHD & FIFO & SIEVE\textsuperscript{!} & 2Q\textsuperscript{!} & 2Q-SIEVE\textsuperscript{!} & ARC\textsuperscript{!} & ARC-SIEVE\textsuperscript{!} & S3-FIFO\textsuperscript{!} & S3-FIFO-SIEVE\textsuperscript{!} \\
\midrule
$U(1,100)$ & 1\% & 10.1 & 13.0 & 11.4 & 5.2 & 11.8 & 9.6 & 11.8 & 10.1 & 12.3 & \cellcolor{gray!18}\textbf{13.4} & 11.8 \\
 & 5\% & 23.6 & 26.8 & 25.9 & 14.9 & 24.0 & 23.2 & 28.8 & 23.5 & \cellcolor{gray!18}\textbf{29.7} & 25.3 & 28.8 \\
 & 10\% & 31.9 & 36.5 & 35.8 & 23.1 & 35.1 & 31.6 & 40.0 & 31.8 & \cellcolor{gray!18}\textbf{40.9} & 33.6 & 40.1 \\
 & 20\% & 44.0 & 47.7 & 48.8 & 35.8 & 50.0 & 43.7 & 53.1 & 43.9 & \cellcolor{gray!18}\textbf{53.6} & 44.8 & 53.1 \\
 & 30\% & 53.4 & 54.5 & 58.1 & 45.9 & 60.3 & 53.3 & 61.2 & 53.4 & \cellcolor{gray!18}\textbf{61.5} & 53.3 & 61.3 \\
$U(1,500)$ & 1\% & 10.2 & 13.2 & 11.5 & 5.2 & 11.8 & 9.8 & 11.9 & 10.2 & 12.5 & \cellcolor{gray!18}\textbf{13.4} & 11.9 \\
 & 5\% & 24.8 & 27.0 & 26.0 & 14.9 & 24.0 & 24.0 & 29.1 & 24.7 & \cellcolor{gray!18}\textbf{30.0} & 25.3 & 29.1 \\
 & 10\% & 33.7 & 36.6 & 36.0 & 23.1 & 35.1 & 33.3 & 40.3 & 33.6 & \cellcolor{gray!18}\textbf{41.1} & 33.6 & 40.3 \\
 & 20\% & 45.6 & 47.8 & 49.0 & 35.8 & 49.9 & 45.3 & 53.2 & 45.5 & \cellcolor{gray!18}\textbf{53.7} & 44.8 & 53.3 \\
 & 30\% & 54.7 & 54.6 & 58.2 & 45.9 & 60.3 & 54.5 & 61.3 & 54.7 & \cellcolor{gray!18}\textbf{61.7} & 53.3 & 61.4 \\
LN(CV=2) & 1\% & 15.9 & \cellcolor{gray!18}\textbf{17.1} & 16.1 & 5.2 & 12.5 & 15.0 & 13.9 & 15.9 & 14.8 & 15.1 & 13.9 \\
 & 5\% & 35.8 & 35.6 & 36.0 & 14.9 & 28.2 & 34.4 & 35.5 & 36.0 & \cellcolor{gray!18}\textbf{37.3} & 27.9 & 35.5 \\
 & 10\% & 48.1 & 46.6 & 48.5 & 23.1 & 41.8 & 47.2 & 49.8 & 49.0 & \cellcolor{gray!18}\textbf{51.9} & 36.6 & 49.9 \\
 & 20\% & 60.7 & 57.0 & 62.8 & 35.7 & 60.6 & 61.7 & 65.0 & 63.8 & \cellcolor{gray!18}\textbf{66.7} & 48.3 & 65.1 \\
 & 30\% & 67.2 & 60.4 & 71.2 & 45.9 & 73.3 & 70.7 & 72.4 & 72.5 & \cellcolor{gray!18}\textbf{73.8} & 57.1 & 72.6 \\
LN(CV=4) & 1\% & \cellcolor{gray!18}\textbf{25.3} & 23.5 & 19.8 & 5.2 & 12.9 & 24.1 & 18.2 & 25.3 & 19.3 & 16.1 & 18.2 \\
 & 5\% & 49.0 & 43.6 & 43.4 & 14.9 & 33.7 & 47.2 & 44.9 & \cellcolor{gray!18}\textbf{49.2} & 47.2 & 29.3 & 44.9 \\
 & 10\% & 60.7 & 54.7 & 56.4 & 23.1 & 49.9 & 59.8 & 61.0 & 61.7 & \cellcolor{gray!18}\textbf{63.2} & 38.1 & 61.0 \\
 & 20\% & 71.4 & 63.8 & 69.5 & 35.7 & 72.1 & 72.5 & 73.8 & 74.2 & \cellcolor{gray!18}\textbf{74.9} & 50.2 & 73.9 \\
 & 30\% & 76.2 & 63.8 & 76.3 & 45.9 & 79.8 & 79.1 & 78.8 & \cellcolor{gray!18}\textbf{80.4} & 79.6 & 59.1 & 78.9 \\
\bottomrule
\end{tabular}%
}
\end{table*}

The synthetic traces show the same pattern. At the $10\%$ cache size,
ARC-SIEVE\textsuperscript{!} gives the highest object hit rate on all four
traces, reaching $51.9\%$ on LN(CV=2) and $63.2\%$ on LN(CV=4). It is
$7\%$--$14\%$ higher than LHD and $11\%$--$16\%$ higher than AdaptSize on these
traces.
At $1\%$ cache size of the two lognormal traces, AdaptSize and LRU-MIN lead
for the same reason as on the META traces: the cache holds only about $10{,}000$
of the $10^6$ objects, so every large object that is admitted displaces many
small popular ones. AdaptSize, which rarely admits large objects, and LRU-MIN,
which evicts them first, strongly protect the small popular objects at this
size; from $5\%$ upward our policies lead.

The size-aware \SIEVE{}-based policies again improve on their parents. Under the
uniform size distributions, 2Q-SIEVE\textsuperscript{!},
ARC-SIEVE\textsuperscript{!}, and S3-FIFO-SIEVE\textsuperscript{!} are
$19\%$--$29\%$ higher than 2Q\textsuperscript{!}, ARC\textsuperscript{!}, and
S3-FIFO\textsuperscript{!} at the $10\%$ cache size. Object sizes vary little
here, so a large part of this gain comes from the \SIEVE{} replacement discipline
itself, which retains reused objects without reordering them, rather than from
the size rule alone. Under the
lognormal distributions, 2Q\textsuperscript{!} and ARC\textsuperscript{!} already
use object size in their victim search, so their \SIEVE{} counterparts gain
$2\%$--$6\%$, while S3-FIFO-SIEVE\textsuperscript{!} is $36\%$--$60\%$ higher than
S3-FIFO\textsuperscript{!}.

\subsection{Runtime Overhead and Byte Hit Rate}
\label{sec:overall-runtime}

\begin{figure}[b]
\centering
\includegraphics[width=\linewidth]{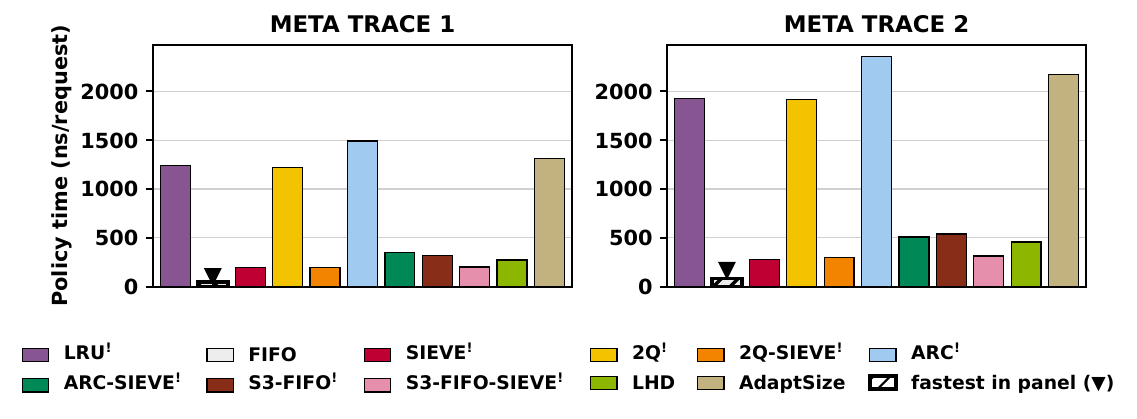}
\caption{Whole-trace replacement-policy overhead at the $10\%$ cache size on
META TRACE 1 and  2. Values are simulator policy time in nanoseconds
per request, so lower bars indicate less CPU and metadata work by the cache
controller. The hatched bar marks the fastest policy in each trace. The numbers
measure policy overhead only, not backend latency or object transfer time. For
FIFO, LHD, and AdaptSize we use the libCacheSim implementations, timed with the
same whole-trace protocol as all other policies.}
\Description{Two bar charts showing policy time per request of our proposed policies and the existing size-aware policies on META TRACE 1 and 2 at the 10 percent cache size.}
\label{fig:section5-overhead}
\end{figure}

Runtime separates two classes of mechanisms. Figure~\ref{fig:section5-overhead}
plots whole-trace policy time at $10\%$ cache size for META TRACE 1 and 2. FIFO is the cheapest baseline. The \SIEVE{}-based hybrids are the next
low-overhead group because they attach the size signal to token assignment or
reinsertion. LHD is slightly more expensive than the \SIEVE{}-based hybrids. In
contrast, LRU\textsuperscript{!}, 2Q\textsuperscript{!}, ARC\textsuperscript{!},
and AdaptSize are substantially more expensive: the first three perform
size-aware victim search, and AdaptSize periodically re-tunes its admission
parameter. Appendix~\ref{app:overall-results} reports the remaining META
timing results at $10\%$ and $30\%$ cache size. As in
Section~\ref{sec:sieve-results}, these times exclude object I/O; in a
deployment, eviction work can be pipelined with the I/O of large objects, so the
differences matter most for caches dominated by small objects.

Byte hit rate measures cache efficiency. It is difficult to discriminate different cache-eviction policies according
to this criterion, see Figure~\ref{fig:app-section5-bhr-meta1} in
Appendix~\ref{app:overall-runtime}. From the $1\%$ cache size upward, FIFO, LHD, and our
size-aware policies have similar byte hit rates. The only exception is AdaptSize, whose byte hit rate is
surprisingly low, at most $12\%$ at every cache size, because its admission rule overly
penalizes large objects. AdaptSize~\cite{BergerAdaptSize17} admits an object with
probability $e^{-\mathrm{size}/c}$, so it almost never caches very large
objects. On the META traces, large objects are requested repeatedly and account for most of the requested bytes, so AdaptSize's byte hit rate can remain low even when its object hit rate is reasonable.

\section{Working-Set Approximation}
\label{sec:wsa}

The empirical results show that, under a cache capacity (byte budget), \SIEVE{}
occupancy depends on both reuse and object size. We model this behavior with a
working-set approximation under the Independent Reference Model (IRM). The main
difference from LRU's characteristic-time approximation is that \SIEVE{} is
controlled by hand examinations rather than by recency movement. A newly admitted
object starts with no retention credit and must be requested before the hand first
examines it. The model therefore uses two characteristic examination times: one from admission
to the first hand examination, and one between subsequent examinations of a
surviving object. Appendix~\ref{app:wsa-details} gives the
full derivation under the IRM for variable-length objects and validates the
WSA against empirical \SIEVE{} replays.

\section{Discussion}
\label{sec:discussion}

\paragraph{Why a running-mean threshold works.}
The (causal) mean-threshold rule uses no future information and adds little state. It
works because very large objects make the mean object size much larger than the
median on the META traces (right-skewed size distribution). 
The rule therefore protects most objects while
withholding protection from the few very large objects that consume
disproportionate cache capacity. The threshold automatically adapts to the workload.
When object sizes vary widely above their mean, the mean moves far above
the median and the benefit is larger. When object-size variance is low, the mean
and median nearly coincide and the benefit fades. The policy does not need a
trace-specific threshold.

\paragraph{Object and byte objectives differ.}
The size rule raises object hit rate because it caches more small reused objects.
A small object and a large object count equally for object hit rate but not for
byte hit rate, and byte traffic on the production traces we used is dominated by very large objects.
Reclaiming space from weakly reused large objects therefore adds many request
hits while barely changing the byte hit rate. This suits caches limited by
request load; a cache limited by bandwidth would need a rule that values bytes
directly.

The direction of the size preference follows from the cache-eviction objective. All other qualities being equal, the
shorter object receives higher retention priority because we evaluate request
hits under a byte budget. An operator optimizing byte hit rate, retrieval costs, or
application value could choose a different priority function
\(\psi(j,\ell_j)\), which assigns a retention priority to object \(j\) of size
\(\ell_j\). For example, a media cache may value retaining a large object
because a miss consumes substantial downstream bandwidth, while a metadata cache
may value retaining many small objects because each hit avoids a backend lookup.
The algorithms would then change the direction or shape of the size rule while
preserving the same causal structure.

\paragraph{Heavier tails give larger gains.}
The gains are largest on the META traces, where the largest $1\%$ of objects
hold more than half of the total object size, and smallest on WIKI TRACE, where
the tail is lighter. The top-$1\%$ share is only a rough guide: META TRACE 3 has
a slightly smaller share than META TRACE 1 but a larger gain. The working-set model gives the same intuition: under a
byte constraint, reducing the occupancy of a few very large objects can free
capacity for many small objects. The WSA expresses this through the byte-weighted capacity condition
$\sum_i L_i h_i=B$ and PASTA~\cite{Wolff89}. A small number of large $L_i$ terms can consume much of the
capacity. Lowering their occupancy can make room for many objects with smaller
$L_i$ but meaningful request probability. When the size tail is weak, the model
has little byte mass to reallocate. The experiments follow the same pattern.

\paragraph{Why the hybridized size-aware \SIEVE{}-based policies perform best.}
In 2Q-SIEVE\textsuperscript{!}, ARC-SIEVE\textsuperscript{!}, and
S3-FIFO-SIEVE\textsuperscript{!}, the segmented parent policy supplies reuse
filtering through its admission path, resident regions, and ghost history. Size-aware \SIEVE{} then decides how much protection each of these
objects receives under the byte budget. Standalone \SIEVE{}\textsuperscript{!}
has no ghost history to keep one-hit objects out,
while the size-aware 2Q\textsuperscript{!}, ARC\textsuperscript{!}, and
S3-FIFO\textsuperscript{!} keep their LRU or FIFO protected regions. Combining the
two gives the highest object hit rates we measured.

For a cache where serving more requests is the primary objective and object
sizes have a heavy tail, 
we therefore recommend adding a running-mean size token to the
\SIEVE{}-managed protected region of a segmented cache. The change is small. It preserves FIFO-style replacement without reordering on
hits and gives consistently strong object hit rates across the measured
workloads.

\paragraph{More retention tokens.}
Two retention tokens do not give the highest hit rate this mechanism can reach.
\SIEVE{}-K8 (Table~\ref{tab:app-sieve-k8}) allows up to
eight tokens per resident object. At the $10\%$ cache size it improves object
hit rate over canonical \SIEVE{} by $8.3\%$--$27.0\%$ on the META traces and by
$27.3\%$--$94.6\%$ on the synthetic traces, and over Dual-Threshold by
$0.8\%$--$2.8\%$ on the META traces and by $16.5\%$--$25.8\%$ on the synthetic
traces. On WIKI TRACE, where the size tail is lighter, the additional tokens
give essentially no further gain over Dual-Threshold. The extra tokens help most
on the synthetic traces, and most of all on the two lognormal traces, which have
the largest relative size dispersion. Wang et al.~\cite{WangSieveClock26} analyze a related design,
a randomized SIEVE/CLOCK variant with multiple access bits per object; combining
multiple bits with size-dependent credit is left for future work.

\paragraph{Limitations.}
Our experiments do not evaluate admission control,
fetch-delay effects, TTL policies~\cite{AtreDelayedHits20,BasuAdaptiveTTL18}, or
latency-aware allocation of cache space across backends~\cite{BergerRobinHood18}.
Cache sizes are percentages of each trace's total object size, so equal
percentages do not mean equal byte capacities across traces. The results also do not claim that object hit rate is always the right objective.
Some deployments are limited by egress bandwidth or storage-device throughput.
Those settings need either a byte-hit objective or a utility function that jointly values
requests and bytes. Our mechanisms target caches whose limiting
resource is request load.

\section{Conclusion}
\label{sec:conclusion}

With variable-length objects, the number of bytes stored in a cache and the
number of requests it serves are no longer proportional. A cache that aims for object hit rate should
therefore not protect a large object and a small object in the same way after
the same weak reuse evidence. A causal running-mean size rule can correct this with little additional 
state. In standalone \SIEVE{}, Dual-Threshold raises object hit rate by up to
$27.7\%$ over canonical \SIEVE{} on the production traces. Hybridized with 2Q, ARC, and
S3-FIFO, the same rule gives our strongest policies. Size-aware 2Q-SIEVE achieves the highest object hit rate for almost all cache sizes across the production traces. It incurs lower policy overhead than the established size-aware policies
LHD, LRU-MIN, and AdaptSize across all measured configurations.
Size-aware ARC-SIEVE achieves the highest object hit rate on the synthetic
traces. The rule keeps \SIEVE{} simple, fits inside existing segmented caches,
and admits a working-set approximation for variable-length objects. The
direction of the size preference follows from the cache-eviction objective, and an operator who
values bytes rather than requests can reverse it within the same framework.

\clearpage
\bibliographystyle{plain}

\begin{thebibliography}{10}

\bibitem{AggarwalSALRU99}
Charu Aggarwal, Joel~L. Wolf, and Philip~S. Yu.
\newblock {Caching on the World Wide Web}.
\newblock {\em IEEE Transactions on Knowledge and Data Engineering}, 11(1):94--107, 1999.

\bibitem{AtreDelayedHits20}
Nirav Atre, Justine Sherry, Weina Wang, and Daniel~S. Berger.
\newblock {Caching with Delayed Hits}.
\newblock In {\em Proceedings of the ACM SIGCOMM Conference}, pages 495--513, 2020.

\bibitem{BansalCAR04}
Sorav Bansal and Dharmendra~S. Modha.
\newblock {CAR: Clock with Adaptive Replacement}.
\newblock In {\em Proceedings of the 3rd USENIX Conference on File and Storage Technologies (FAST)}, pages 187--200, 2004.

\bibitem{BasuAdaptiveTTL18}
Soumya Basu, Aditya Sundarrajan, Javad Ghaderi, Sanjay Shakkottai, and Ramesh Sitaraman.
\newblock {Adaptive TTL-Based Caching for Content Delivery}.
\newblock {\em IEEE/ACM Transactions on Networking}, 26(3):1063--1077, 2018.

\bibitem{BeckmannLHD18}
Nathan Beckmann, Haoxian Chen, and Asaf Cidon.
\newblock {LHD: Improving Cache Hit Rate by Maximizing Hit Density}.
\newblock In {\em Proceedings of the 15th USENIX Symposium on Networked Systems Design and Implementation (NSDI)}, pages 389--403, 2018.

\bibitem{Belady66}
L.~A. Belady.
\newblock {A Study of Replacement Algorithms for a Virtual-Storage Computer}.
\newblock {\em IBM Systems Journal}, 5(2):78--101, 1966.

\bibitem{BergCacheLib20}
Benjamin Berg, Daniel~S. Berger, Sara McAllister, Isaac Grosof, Sathya Gunasekar, Jimmy Lu, Michael Uhlar, Jim Carrig, Nathan Beckmann, Mor Harchol-Balter, and Gregory~R. Ganger.
\newblock {The CacheLib Caching Engine: Design and Experiences at Scale}.
\newblock In {\em Proceedings of the 14th USENIX Symposium on Operating Systems Design and Implementation (OSDI)}, pages 753--768, 2020.

\bibitem{BergerFOO18}
Daniel~S. Berger, Nathan Beckmann, and Mor Harchol-Balter.
\newblock {Practical Bounds on Optimal Caching with Variable Object Sizes}.
\newblock In {\em Proceedings of the ACM on Measurement and Analysis of Computing Systems (SIGMETRICS)}, volume~2, pages 32:1--32:38, 2018.

\bibitem{BergerRobinHood18}
Daniel~S. Berger, Benjamin Berg, Timothy Zhu, Siddhartha Sen, and Mor Harchol-Balter.
\newblock {RobinHood: Tail Latency Aware Caching -- Dynamic Reallocation from Cache-Rich to Cache-Poor}.
\newblock In {\em Proceedings of the 13th USENIX Symposium on Operating Systems Design and Implementation (OSDI)}, pages 195--212, 2018.

\bibitem{BergerAdaptSize17}
Daniel~S. Berger, Ramesh~K. Sitaraman, and Mor Harchol-Balter.
\newblock {AdaptSize: Orchestrating the Hot Object Memory Cache in a Content Delivery Network}.
\newblock In {\em Proceedings of the 14th USENIX Symposium on Networked Systems Design and Implementation (NSDI)}, pages 483--498, 2017.

\bibitem{BhattacharjeeRegret20}
Rajarshi Bhattacharjee, Subhankar Banerjee, and Abhishek Sinha.
\newblock {Fundamental Limits on the Regret of Online Network-Caching}.
\newblock {\em Proceedings of the ACM on Measurement and Analysis of Computing Systems}, 4(2):1--31, 2020.

\bibitem{CaoIrani97}
Pei Cao and Sandy Irani.
\newblock {Cost-Aware WWW Proxy Caching Algorithms}.
\newblock In {\em Proceedings of the USENIX Symposium on Internet Technologies and Systems (USITS)}, pages 193--206, 1997.

\bibitem{CTZ02}
Hao Che, Ye~Tung, and Zhijun Wang.
\newblock {Hierarchical Web Caching Systems: Modeling, Design and Experimental Results}.
\newblock {\em IEEE Journal on Selected Areas in Communications}, 20(7):1305--1314, 2002.

\bibitem{ChengLRUSP00}
Kai Cheng and Yahiko Kambayashi.
\newblock {LRU-SP: A Size-Adjusted and Popularity-Aware LRU Replacement Algorithm for Web Caching}.
\newblock In {\em Proceedings of the 24th Annual International Computer Software and Applications Conference (COMPSAC)}, pages 48--53, 2000.

\bibitem{CidonCliffhanger16}
Asaf Cidon, Assaf Eisenman, Mohammad Alizadeh, and Sachin Katti.
\newblock {Cliffhanger: Scaling Performance Cliffs in Web Memory Caches}.
\newblock In {\em Proceedings of the 13th USENIX Symposium on Networked Systems Design and Implementation (NSDI)}, pages 379--392, 2016.

\bibitem{DehghanUtility16}
Mostafa Dehghan, Laurent Massoulie, Don Towsley, Daniel~S. Menasche, and Y.~C. Tay.
\newblock {A Utility Optimization Approach to Network Cache Design}.
\newblock In {\em Proceedings of IEEE INFOCOM}, pages 1--9, 2016.

\bibitem{DeS72}
Peter~J. Denning and Stuart~C. Schwartz.
\newblock {Properties of the Working-Set Model}.
\newblock {\em Communications of the ACM}, 15(3):191--198, 1972.

\bibitem{DongLockFree25}
Chao Dong, Fang Wang, Hong Jiang, and Dan Feng.
\newblock {Using Lock-Free Design for Throughput-Optimized Cache Eviction}.
\newblock {\em Proceedings of the ACM on Measurement and Analysis of Computing Systems}, 9(2):1--28, 2025.

\bibitem{FerragutRP16}
Andres Ferragut, Ismael Rodr\'{i}guez, and Fernando Paganini.
\newblock {Optimizing TTL Caches under Heavy-Tailed Demands}.
\newblock In {\em Proceedings of the ACM SIGMETRICS International Conference on Measurement and Modeling of Computer Systems}, pages 101--112, 2016.

\bibitem{FrickerRoberts12}
Christine Fricker, Philippe Robert, and James Roberts.
\newblock {A Versatile and Accurate Approximation for LRU Cache Performance}.
\newblock In {\em Proceedings of the 24th International Teletraffic Congress (ITC)}, pages 8:1--8:8, 2012.

\bibitem{GarettoLM16}
Michele Garetto, Emilio Leonardi, and Valentina Martina.
\newblock {A Unified Approach to the Performance Analysis of Caching Systems}.
\newblock {\em ACM Transactions on Modeling and Performance Evaluation of Computing Systems}, 1(3):12:1--12:28, 2016.

\bibitem{GastVanHoudt15}
Nicolas Gast and Benny Van~Houdt.
\newblock {Transient and Steady-State Regime of a Family of List-Based Cache Replacement Algorithms}.
\newblock In {\em Proceedings of the ACM SIGMETRICS International Conference on Measurement and Modeling of Computer Systems}, pages 123--136, 2015.

\bibitem{HanScaleOPT24}
Hyungseok Han, Sangjin Lee, and Yongseok Son.
\newblock {ScaleOPT: A Scalable Optimal Page Replacement Policy Simulator}.
\newblock {\em Proceedings of the ACM on Measurement and Analysis of Computing Systems}, 8(3):1--25, 2024.

\bibitem{JiangClockPro05}
Song Jiang, Feng Chen, and Xiaodong Zhang.
\newblock {CLOCK-Pro: An Effective Improvement of the CLOCK Replacement}.
\newblock In {\em Proceedings of the USENIX Annual Technical Conference (ATC)}, pages 323--336, 2005.

\bibitem{JiangLIRS02}
Song Jiang and Xiaodong Zhang.
\newblock {LIRS: An Efficient Low Inter-Reference Recency Set Replacement Policy to Improve Buffer Cache Performance}.
\newblock In {\em Proceedings of the ACM SIGMETRICS International Conference on Measurement and Modeling of Computer Systems}, pages 31--42, 2002.

\bibitem{JohnsonShasha94}
Theodore Johnson and Dennis Shasha.
\newblock {2Q: A Low Overhead High Performance Buffer Management Replacement Algorithm}.
\newblock In {\em Proceedings of the 20th International Conference on Very Large Data Bases}, VLDB '94, pages 439--450, San Francisco, CA, USA, 1994. Morgan Kaufmann Publishers Inc.

\bibitem{LiOnlineCachingNetworks21}
Yuanyuan Li, Tareq Si~Salem, Giovanni Neglia, and Stratis Ioannidis.
\newblock {Online Caching Networks with Adversarial Guarantees}.
\newblock {\em Proceedings of the ACM on Measurement and Analysis of Computing Systems}, 5(3):1--39, 2021.

\bibitem{MartinaGaretto14}
Valentina Martina, Michele Garetto, and Emilio Leonardi.
\newblock {A Unified Approach to the Performance Analysis of Caching Systems}.
\newblock In {\em Proceedings of IEEE INFOCOM}, pages 2040--2048, 2014.

\bibitem{MegiddoModha03}
Nimrod Megiddo and Dharmendra~S. Modha.
\newblock {ARC: A Self-Tuning, Low Overhead Replacement Cache}.
\newblock {\em Proceedings of the USENIX Conference on File and Storage Technologies}, pages 115--130, 2003.

\bibitem{ProductionTraces}
{Meta Platforms, Inc.}
\newblock {Evaluating SSD Hardware for Facebook Workloads}.
\newblock \url{https://cachelib.org/docs/Cache_Library_User_Guides/Cachebench_FB_HW_eval/}, 2026.
\newblock CacheLib documentation.

\bibitem{MhaisenOptimistic22}
Naram Mhaisen, Abhishek Sinha, Georgios Paschos, and George Iosifidis.
\newblock {Optimistic No-regret Algorithms for Discrete Caching}.
\newblock {\em Proceedings of the ACM on Measurement and Analysis of Computing Systems}, 6(3):1--28, 2022.

\bibitem{NegliaCM18}
Giovanni Neglia, Damiano Carra, and Pietro Michiardi.
\newblock {Cache Policies for Linear Utility Maximization}.
\newblock {\em IEEE/ACM Transactions on Networking}, 26(1):302--313, 2018.

\bibitem{NygrenAkamai10}
Erik Nygren, Ramesh~K. Sitaraman, and Jennifer Sun.
\newblock {The Akamai Network: A Platform for High-Performance Internet Applications}.
\newblock {\em ACM SIGOPS Operating Systems Review}, 44(3):2--19, 2010.

\bibitem{ONeilLRUK93}
Elizabeth~J. O'Neil, Patrick~E. O'Neil, and Gerhard Weikum.
\newblock {The LRU-K Page Replacement Algorithm for Database Disk Buffering}.
\newblock In {\em Proceedings of the ACM SIGMOD International Conference on Management of Data}, pages 297--306, 1993.

\bibitem{QiuHitRatioThroughput26}
Ziyue Qiu, Juncheng Yang, and Mor Harchol-Balter.
\newblock {Why Increasing the Hit Ratio Can Hurt Cache Throughput}.
\newblock {\em Queueing Systems}, 110(2):28, 2026.

\bibitem{QuanMultiFlowLRU19}
Guocong Quan, Jian Tan, Atilla Eryilmaz, and Ness Shroff.
\newblock {A New Flexible Multi-flow LRU Cache Management Paradigm for Minimizing Misses}.
\newblock {\em Proceedings of the ACM on Measurement and Analysis of Computing Systems}, 3(2):1--30, 2019.

\bibitem{SongLRB20}
Zhenyu Song, Daniel~S. Berger, Kai Li, and Wyatt Lloyd.
\newblock {Learning Relaxed Belady for Content Distribution Network Caching}.
\newblock In {\em Proceedings of the 17th USENIX Symposium on Networked Systems Design and Implementation (NSDI)}, pages 529--544, 2020.

\bibitem{SundarrajanCoNEXT17}
Aditya Sundarrajan, Mingdong Feng, Mangesh Kasbekar, and Ramesh~K. Sitaraman.
\newblock {Footprint Descriptors: Theory and Practice of Cache Provisioning in a Global CDN}.
\newblock In {\em Proceedings of the 13th International Conference on Emerging Networking EXperiments and Technologies (CoNEXT)}, pages 55--67, 2017.

\bibitem{WangSieveClock26}
Yirong Wang, Peter Desnoyers, and Benny Van~Houdt.
\newblock {Performance Analysis of the Randomized SIEVE/CLOCK Cache Replacement Algorithm}.
\newblock {\em Proceedings of the ACM on Measurement and Analysis of Computing Systems}, 10(2):1--26, 2026.

\bibitem{WilliamsLRUMIN96}
Stephen Williams, Marc Abrams, Charles~R. Standridge, Ghaleb Abdulla, and Edward~A. Fox.
\newblock {Removal Policies in Network Caches for World-Wide Web Documents}.
\newblock In {\em Proceedings of the ACM SIGCOMM Conference}, pages 293--305, 1996.

\bibitem{Wolff89}
Ronald~W. Wolff.
\newblock {Poisson Arrivals See Time Averages}.
\newblock {\em Operations Research}, 30(2):223--231, 1982.

\bibitem{WolffBook89}
Ronald~W. Wolff.
\newblock {\em {Stochastic Modeling and the Theory of Queues}}.
\newblock Prentice Hall, Englewood Cliffs, NJ, 1989.

\bibitem{YangGLCache23}
Juncheng Yang, Ziming Mao, Yao Yue, and K.~V. Rashmi.
\newblock {GL-Cache: Group-Level Learning for Efficient and High-Performance Caching}.
\newblock In {\em Proceedings of the 21st USENIX Conference on File and Storage Technologies (FAST)}, pages 115--134, 2023.

\bibitem{YangTwitter20}
Juncheng Yang, Yao Yue, and K.~V. Rashmi.
\newblock {A Large-Scale Analysis of Hundreds of In-Memory Cache Clusters at Twitter}.
\newblock In {\em Proceedings of the 14th USENIX Symposium on Operating Systems Design and Implementation (OSDI)}, pages 191--208, 2020.

\bibitem{YZQ23}
Juncheng Yang, Yazhuo Zhang, Ziyue Qiu, Yao Yue, and Rashmi Vinayak.
\newblock {FIFO Queues Are All You Need for Cache Eviction}.
\newblock In {\em Proceedings of the 29th Symposium on Operating Systems Principles}, SOSP '23, pages 130--149, New York, NY, USA, 2023. Association for Computing Machinery.

\bibitem{Young02}
Neal~E. Young.
\newblock {On-Line File Caching}.
\newblock {\em Algorithmica}, 33(3):371--383, 2002.

\bibitem{ZYY24}
Yazhuo Zhang, Juncheng Yang, Yao Yue, Ymir Vigfusson, and K.~V. Rashmi.
\newblock {SIEVE is Simpler than LRU: an Efficient Turn-Key Eviction Algorithm for Web Caches}.
\newblock In {\em Proceedings of the 21st USENIX Symposium on Networked Systems Design and Implementation (NSDI '24)}, pages 1229--1246. USENIX Association, 2024.

\end{thebibliography}

\clearpage

\setcounter{section}{0}
\renewcommand{\thesection}{\Alph{section}}
\renewcommand{\theHsection}{appendix.\Alph{section}}

\section{Background on Existing Policies}
\label{app:background}

This appendix gives the descriptions of the existing policies that
Section~\ref{sec:background} summarizes: Appendices~\ref{app:canonical-sieve},
\ref{app:established}, and~\ref{app:segmented} supplement
Sections~\ref{sec:canonical-sieve}, \ref{sec:lru-components},
and~\ref{sec:blended-construction}.

\subsection{Canonical \SIEVE{}}
\label{app:canonical-sieve}

\SIEVE{} keeps resident objects in insertion order and associates one visited
bit with each resident object~\cite{ZYY24}. Let
\(Q_t=(q_1,q_2,\ldots,q_m)\) denote the resident ordering at time \(t\), from
oldest to newest, and let \(v_x\in\{0,1\}\) be the visited bit of resident
object \(x\). A miss inserts the requested object at the newest end with
\(v_x=0\). A hit changes only the bit,
\begin{equation}
  v_x \gets 1,
\label{eq:sieve-hit}
\end{equation}
and leaves the resident order unchanged. Thus the hit path is constant time and
does not perform an LRU-style move.

When additional bytes are needed, an eviction hand scans residents from older to
newer entries. If the hand finds an object with \(v_x=0\), that object is
removed. If it finds \(v_x=1\), the bit is cleared and the hand advances:
\begin{equation}
  v_x \gets 0 \qquad \text{and the hand advances.}
\label{eq:sieve-clear}
\end{equation}
The first zero-bit object encountered by the scan is the victim. Since newly
inserted objects have \(v_x=0\), an object admitted after a miss receives no
retention credit unless it is requested again before the hand reaches it.

\begin{algorithm}[H]
\caption{Canonical \SIEVE{}}
\label{alg:canonical-sieve}
\vspace{2pt}\hrule\vspace{4pt}
\begin{tabbing}
\quad\=\quad\=\kill
\textbf{State:} insertion-ordered resident queue; hand; bit \(v_x\in\{0,1\}\).\\[1pt]
\textbf{On hit to \(x\):} \(v_x\gets 1\).\\[1pt]
\textbf{On miss for admissible \(x\):} insert \(x\) at the newest end with \(v_x\gets0\).\\[2pt]
\textbf{While the cache exceeds capacity:}\\
\> \(y\gets\) object currently at the hand\\
\> \textbf{if} \(v_y=0\) \textbf{then} evict \(y\)\\
\> \textbf{else} \(v_y\gets0\) and advance the hand
\end{tabbing}
\vspace{2pt}\hrule
\end{algorithm}

This deferred use of the hit signal is the key property we exploit. \SIEVE{}
records reuse immediately, but the replacement consequence of that reuse is
resolved only when the hand later examines the object. Size can therefore alter
retention at the hand without changing the constant-time hit path or requiring a
new ordering of the cache.

\subsection{Established Size-Aware Policies}
\label{app:established}

\paragraph{LRU-MIN}
LRU-MIN was introduced for Web proxy caches containing variable-sized documents.
When an incoming object of size \(S\) must be admitted, the policy first asks
whether any resident object of size at least \(S\) can be removed. If so, it
chooses the least-recently-used object among that eligible set. If not, the
threshold is relaxed to \(S/2\), then \(S/4\), and so on until a victim can be
selected. LRU-MIN therefore preserves the recency principle within each eligible
size class, but it uses the incoming object's size to avoid freeing space
through a long sequence of small evictions. In Section~\ref{sec:overall-results},
we report this LRU-family size-aware baseline as LRU\textsuperscript{!}.

\paragraph{AdaptSize}
AdaptSize is a size-aware admission policy for content-delivery caches. On a
miss for object \(i\), with size \(s_i\), the object is admitted with probability
\begin{equation}
    p_i(c) = \exp(-s_i/c),
    \label{eq:adaptsize-admission}
\end{equation}
where \(c\) is the admission parameter. Small objects therefore have high
admission probability, while large objects are admitted only when repeated
requests make them likely to pass the probabilistic filter. Objects that are
admitted are managed by an LRU-like eviction rule; objects rejected by admission
are served but not stored.

The main technical problem in AdaptSize is choosing \(c\). For a recent request
window, let \(r_i\) be the measured request rate of object \(i\), and let
\(x_i(c,\mu_c)\) denote the stationary probability that object \(i\) is resident
under AdaptSize's per-object Markov-chain model. The common pushdown rate
\(\mu_c\) is chosen so that the expected resident byte occupancy equals the cache
capacity \(K\):
\begin{equation}
    \sum_i s_i x_i(c,\mu_c) = K .
    \label{eq:adaptsize-capacity}
\end{equation}
AdaptSize then predicts the object hit rate for a candidate \(c\) as
\begin{equation}
    \widehat{H}_{\mathrm{obj}}(c)
    = \frac{\sum_i r_i x_i(c,\mu_c)}{\sum_i r_i},
    \label{eq:adaptsize-ohr}
\end{equation}
and periodically selects the \(c\) that maximizes this predicted hit rate. Thus
AdaptSize is not simply a fixed size cutoff: it is a probabilistic admission
rule whose size bias is tuned online from recent workload statistics.

\paragraph{LHD}
LHD, or least hit density, is an eviction policy that ranks resident objects
by predicted hits per byte-time consumed. Consider an object of size \(s_i\), age
\(a\), and equivalence class \(c\), where the class can encode features such as
reference count or application identity. Let \(H\) be the random hit age and
\(L\) the random lifetime age, where a lifetime ends either in a hit or in an
eviction. LHD estimates the object's remaining hit density as
\begin{equation}
    D_i(a,c) =
    \frac{\sum_{x\ge 1} \Pr[H=a+x\mid c]}
    {s_i \sum_{x\ge 1} x\Pr[L=a+x\mid c]} .
    \label{eq:lhd-density}
\end{equation}
The numerator estimates the probability that the object will hit in the future,
conditioned on having survived to age \(a\). The denominator estimates the
remaining byte-time the object will consume. LHD updates the hit- and
lifetime-age distributions online and evicts the resident object with the
smallest \(D_i(a,c)\). Unlike LRU-MIN, which searches for size-compatible LRU
victims, or AdaptSize, which filters admissions, LHD makes size part of an
explicit utility estimate for every eviction candidate.

\subsection{Segmented and Blended Cache Policies}
\label{app:segmented}

Many replacement policies separate newly admitted objects from objects that
have accumulated stronger evidence of reuse. We refer to these regions as
\emph{cold} and \emph{hot}, respectively. Cold and hot describe their functional
roles rather than a particular data structure: a region may be managed by FIFO,
LRU, \SIEVE{}, or another replacement primitive.

This separation provides a reuse filter. A newly observed object is initially
given limited protection; only after satisfying the parent policy's promotion
rule does it enter the hot region. Several policies also retain ghost histories
for recently evicted objects. Because ghosts contain metadata rather than
payload, a ghost hit is a miss, but it provides evidence that the object
returned after eviction. The parent policies differ primarily in how this
evidence is collected and how it changes admission, promotion, or the division
of resident capacity. How to divide an LRU cache among request classes has also
been analyzed directly~\cite{QuanMultiFlowLRU19}.

\paragraph{2Q}
2Q separates recent arrivals from objects that demonstrate reuse using three
structures~\cite{JohnsonShasha94}: a resident FIFO queue for recently admitted
objects, a nonresident ghost history for objects evicted from that queue, and
an LRU queue for objects receiving longer-term protection. A previously unseen
object enters the FIFO region. A hit while the object is still resident there
does not immediately move it to the protected queue. If the object is evicted
and subsequently requested while its identifier remains in the ghost history,
the returning object is admitted to the protected LRU queue. The ghost
therefore distinguishes a second access separated by eviction from a short
burst of references while the object is still resident.

\paragraph{ARC}
ARC maintains two resident LRU lists and two corresponding ghost
lists~\cite{MegiddoModha03}. One resident list represents objects seen
recently, while the other represents objects with stronger repeated-reference
evidence. Their ghost histories record objects recently evicted from the two
resident lists. A hit in the recency ghost history indicates that the recency
side was too small; a hit in the frequency ghost history provides evidence in
the opposite direction. ARC uses these events to adapt the target division
between the two resident lists online. Thus, unlike a cache with a fixed
cold/hot partition, ARC changes its preferred balance between recency and
frequency as the request stream evolves.

\paragraph{S3-FIFO}
S3-FIFO uses FIFO structures rather than LRU lists~\cite{YZQ23}. A small
resident FIFO region filters newly admitted objects, a larger resident FIFO
region protects objects with stronger reuse evidence, and a FIFO ghost queue
records recently evicted identifiers. Hits update a small reuse counter without
changing FIFO order. When an object reaches the eviction end of the small
region, the counter determines whether the object is promoted or evicted to
the ghost queue. A subsequent request to a ghosted object admits the returning
object with stronger protection. In the main FIFO region, counter-based
reinsertion gives reused objects additional opportunities to remain resident
without requiring LRU-style movement on every hit.

\paragraph{\SIEVE{}-based blended variants.}
\SIEVE{} can also be used as a replacement primitive inside an existing
segmented policy~\cite{ZYY24}. We call such constructions \emph{blended}
policies: the parent policy retains its admission path, ghost histories,
partitioning rule, and promotion semantics, but one resident replacement
component is managed by \SIEVE{} instead of the parent's original primitive.

In 2Q-SIEVE, the FIFO admission region and ghost-hit rule of 2Q are unchanged,
while the protected LRU queue is replaced by \SIEVE{}. ARC-SIEVE retains ARC's
ghost histories and adaptive target split but uses \SIEVE{} for replacement in
its protected, frequency-oriented resident region. S3-FIFO-SIEVE retains
S3-FIFO's small FIFO filter and ghost behavior while managing its protected
region using \SIEVE{}. This construction is useful experimentally because the
parent policy's reuse evidence remains fixed while the local eviction
mechanism changes.

Table~\ref{tab:policy-roles} summarizes the roles of the principal baseline and
segmented policies.

\begin{table}[H]
\centering
\caption{Resident organization and reuse signals of the principal policies.
Ghost entries contain metadata only and do not count toward resident capacity
\(B\).}
\label{tab:policy-roles}
\small
\setlength{\tabcolsep}{3.5pt}
\renewcommand{\arraystretch}{1.10}
\begin{tabular}{p{0.20\linewidth} p{0.34\linewidth} p{0.36\linewidth}}
\toprule
Policy & Resident organization & Reuse or adaptation signal \\
\midrule
FIFO
& One FIFO queue
& Hits do not change order \\

LRU
& One LRU list
& Hits refresh recency \\

\SIEVE{}
& Insertion-ordered queue
& Visited bit gives lazy protection \\

2Q
& FIFO cold, LRU hot
& Cold ghost return admits to hot \\

ARC
& Two resident LRU lists
& Ghost hits adapt recency/frequency split \\

S3-FIFO
& FIFO cold, FIFO hot
& Counters and ghost returns identify reuse \\

2Q-SIEVE
& FIFO cold, \SIEVE{} hot
& 2Q admission with \SIEVE{} hot replacement \\

ARC-SIEVE
& ARC resident regions with \SIEVE{} hot
& ARC adaptation with \SIEVE{} hot replacement \\

S3-FIFO-SIEVE
& FIFO cold, \SIEVE{} hot
& S3-FIFO admission with \SIEVE{} hot replacement \\
\bottomrule
\end{tabular}
\end{table}

\section{Additional Trace Characterization}
\label{app:trace-characterization}

\subsection{Production Traces}
\label{app:production-traces}

\begin{figure}[H]
\centering
\begin{minipage}{0.49\textwidth}
\centering
\includegraphics[width=\linewidth]{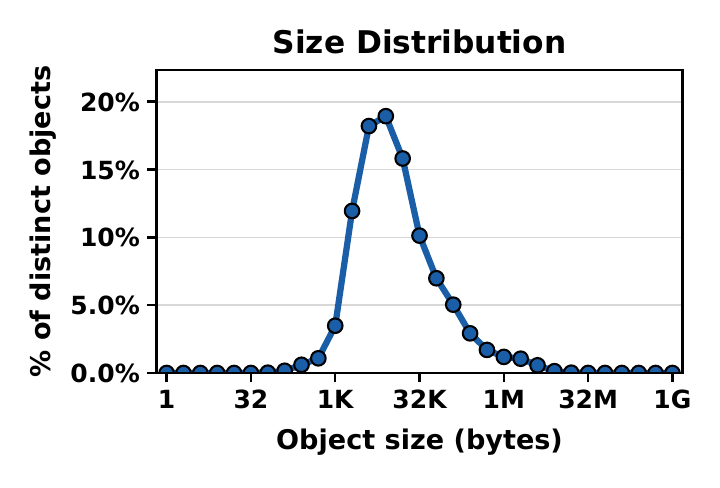}
{\small (a) WIKI TRACE: size distribution}
\end{minipage}\hfill
\begin{minipage}{0.49\textwidth}
\centering
\includegraphics[width=\linewidth]{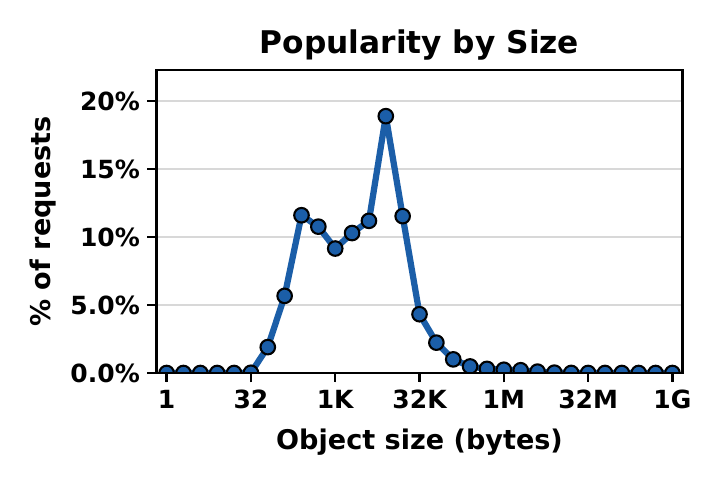}
{\small (b) WIKI TRACE: popularity by size}
\end{minipage}
\caption{WIKI TRACE production-trace characterization. Panel (a) shows the
share of distinct objects in each object-size range, and panel (b) shows the
request popularity contributed by objects in the same size ranges. WIKI TRACE
is still heterogeneous, but its popularity is more concentrated in smaller size
ranges than in the META traces, so the expected benefit of size-aware eviction
is more moderate.}
\Description{Two panels showing WIKI TRACE object-size distribution and
popularity by object size.}
\label{fig:app-wiki-size-pop}
\end{figure}

\begin{figure}[H]
\centering
\begin{minipage}{0.49\textwidth}
\centering
\includegraphics[width=\linewidth]{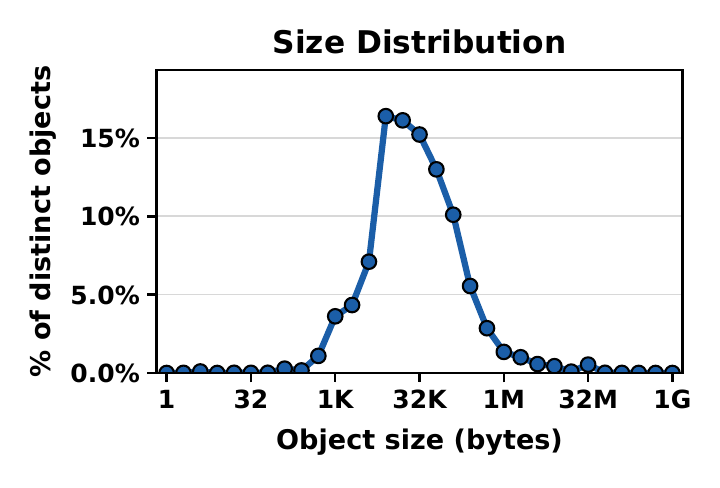}
{\small (a) META TRACE 2: size distribution}
\end{minipage}\hfill
\begin{minipage}{0.49\textwidth}
\centering
\includegraphics[width=\linewidth]{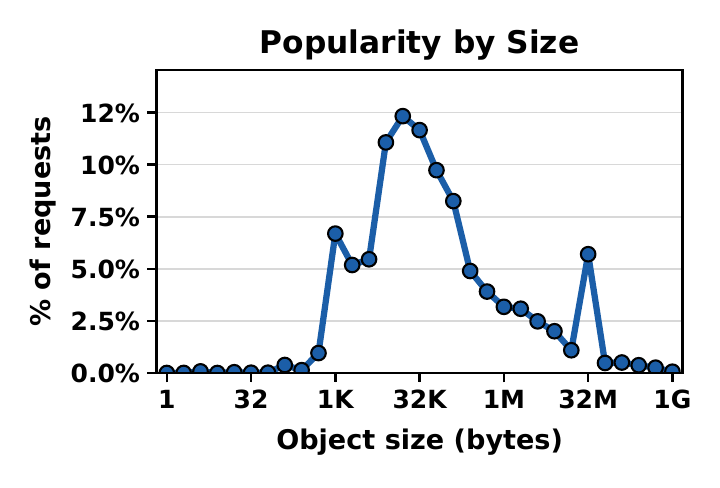}
{\small (b) META TRACE 2: popularity by size}
\end{minipage}
\caption{META TRACE 2 production-trace characterization. Panel (a) shows the
share of distinct objects in each object-size range, and panel (b) shows the
request popularity contributed by objects in the same size ranges. The trace has
a heavy size tail and request mass spread beyond the smallest objects, which
creates byte pressure for size-oblivious replacement policies.}
\Description{Two panels showing META TRACE 2 object-size distribution and
popularity by object size.}
\label{fig:app-meta2-size-pop}
\end{figure}

\begin{figure}[H]
\centering
\begin{minipage}{0.49\textwidth}
\centering
\includegraphics[width=\linewidth]{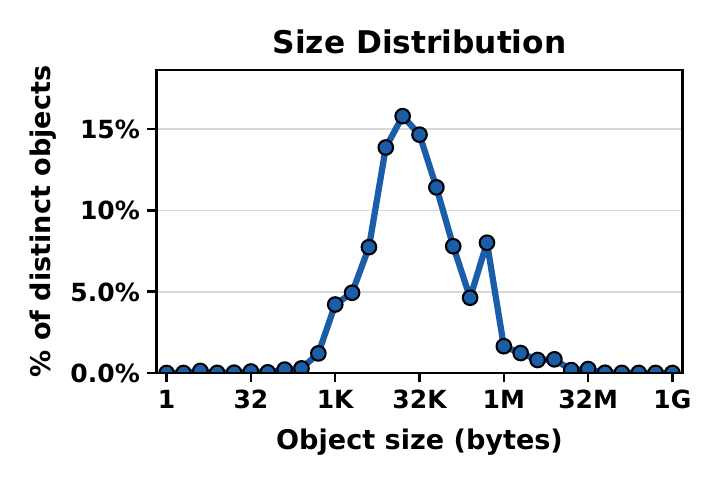}
{\small (a) META TRACE 3: size distribution}
\end{minipage}\hfill
\begin{minipage}{0.49\textwidth}
\centering
\includegraphics[width=\linewidth]{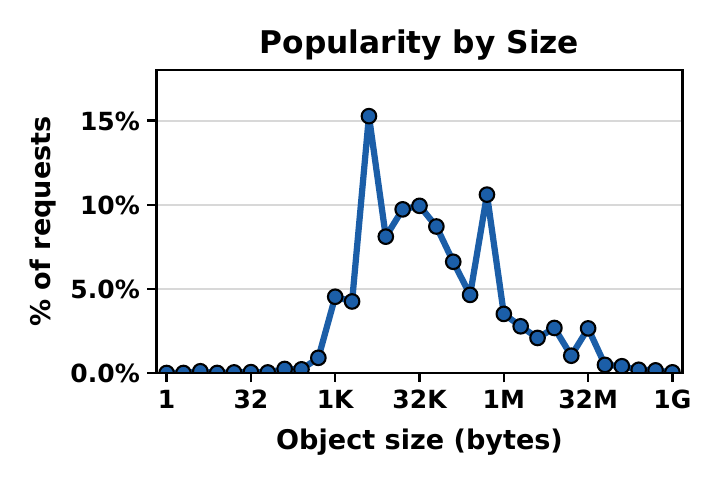}
{\small (b) META TRACE 3: popularity by size}
\end{minipage}
\caption{META TRACE 3 production-trace characterization. Panel (a) reports the
object-size distribution over distinct objects, and panel (b) reports the
request popularity associated with each size range. As with the other META
traces, substantial heterogeneity remains among requested objects, so equal
retention credit for large and small objects can waste scarce byte capacity.}
\Description{Two panels showing META TRACE 3 object-size distribution and
popularity by object size.}
\label{fig:app-meta3-size-pop}
\end{figure}

This subsection supplements Section~\ref{sec:production-traces}.
Figures~\ref{fig:app-wiki-size-pop}--\ref{fig:app-meta3-size-pop} provide the
same size and popularity views for WIKI TRACE, META TRACE 2 and  3
that Figure~\ref{fig:size-dist} shows for META TRACE 1. In both additional META
traces, popularity is spread across multiple object-size ranges rather than
being isolated in a single size bucket. This is the regime where byte-capacity
allocation matters: retaining a large object consumes space that could otherwise
hold many smaller objects with comparable request evidence.

\subsection{Synthetic Traces}
\label{app:synthetic-traces}

This subsection supplements Section~\ref{sec:synthetic-traces}.
Table~\ref{tab:synthetic-workloads} lists the size statistics of the four
synthetic traces.

\begin{table}[H]
\centering
\caption{Synthetic traces. $U(a,b)$ draws object sizes
uniformly from $a$ to $b$ bytes, and LN(CV=$c$) draws them from a lognormal
distribution. $\mu_L$ is the mean object size, $\sigma_L$ is the standard
deviation of object size, and CV$\,=\sigma_L/\mu_L$ is the coefficient of
variation. All four traces use $N=10^6$ objects, $Q=10^7$ requests, and Zipf
parameter $\alpha=0.6$.}
\label{tab:synthetic-workloads}
\small
\begin{tabular}{lrrr}
\toprule
Trace & Mean size $\mu_L$ (bytes) & Std.\ deviation $\sigma_L$ (bytes) & CV $=\sigma_L/\mu_L$ \\
\midrule
$U(1,100)$ & 50.5 & 28.9 & 0.57 \\
$U(1,500)$ & 250.4 & 144.4 & 0.58 \\
LN(CV=2) & 50.0 & 99.6 & 1.99 \\
LN(CV=4) & 51.4 & 199.0 & 3.87 \\
\bottomrule
\end{tabular}

\end{table}

\section{Full Size-Aware \SIEVE{} Results}
\label{app:sieve}

This appendix supplements Section~\ref{sec:sieve-results}; its subsections follow
the subsections of that section.

\subsection{Production Traces}
\label{app:sieve-production}

This subsection supplements Section~\ref{sec:sieve-production}.
Tables~\ref{tab:app-sieve-meta_trace1}--\ref{tab:app-sieve-wiki_2016u} report the
full production object-hit-rate sweep for canonical \SIEVE{}, MT-SIEVE, and
DT-SIEVE.

\begin{table}[H]
\centering
\caption{META TRACE 1: object hit rate (\%) for canonical \SIEVE{}, MT-SIEVE (Mean-Threshold \SIEVE{}), and DT-SIEVE (Dual-Threshold \SIEVE{}). Best per row in bold.}
\label{tab:app-sieve-meta_trace1}
\small
\begin{tabular}{lrrr}
\toprule
Cache & Canonical & MT-SIEVE & DT-SIEVE \\
\midrule
0.01\% & 49.93 & 51.11 & \cellcolor{gray!18}\textbf{53.05} \\
0.1\% & 55.19 & 56.70 & \cellcolor{gray!18}\textbf{59.71} \\
1\% & 60.79 & 62.42 & \cellcolor{gray!18}\textbf{65.57} \\
5\% & 64.82 & 66.74 & \cellcolor{gray!18}\textbf{70.24} \\
10\% & 66.86 & 68.87 & \cellcolor{gray!18}\textbf{71.87} \\
20\% & 69.00 & 70.97 & \cellcolor{gray!18}\textbf{72.80} \\
30\% & 70.17 & 71.93 & \cellcolor{gray!18}\textbf{72.90} \\
\bottomrule
\end{tabular}
\end{table}

\begin{table}[H]
\centering
\caption{META TRACE 2: object hit rate (\%) for canonical \SIEVE{}, MT-SIEVE (Mean-Threshold \SIEVE{}), and DT-SIEVE (Dual-Threshold \SIEVE{}). Best per row in bold.}
\label{tab:app-sieve-meta_trace2}
\small
\begin{tabular}{lrrr}
\toprule
Cache & Canonical & MT-SIEVE & DT-SIEVE \\
\midrule
0.01\% & 24.01 & 25.22 & \cellcolor{gray!18}\textbf{26.55} \\
0.1\% & 30.61 & 32.20 & \cellcolor{gray!18}\textbf{35.94} \\
1\% & 37.34 & 40.07 & \cellcolor{gray!18}\textbf{46.40} \\
5\% & 44.44 & 48.15 & \cellcolor{gray!18}\textbf{56.77} \\
10\% & 48.52 & 52.71 & \cellcolor{gray!18}\textbf{60.57} \\
20\% & 53.48 & 57.81 & \cellcolor{gray!18}\textbf{62.62} \\
30\% & 56.76 & 60.43 & \cellcolor{gray!18}\textbf{62.94} \\
\bottomrule
\end{tabular}
\end{table}

\begin{table}[H]
\centering
\caption{META TRACE 3: object hit rate (\%) for canonical \SIEVE{}, MT-SIEVE (Mean-Threshold \SIEVE{}), and DT-SIEVE (Dual-Threshold \SIEVE{}). Best per row in bold.}
\label{tab:app-sieve-meta_trace3}
\small
\begin{tabular}{lrrr}
\toprule
Cache & Canonical & MT-SIEVE & DT-SIEVE \\
\midrule
0.01\% & 30.29 & 31.31 & \cellcolor{gray!18}\textbf{32.75} \\
0.1\% & 38.02 & 39.43 & \cellcolor{gray!18}\textbf{42.20} \\
1\% & 44.84 & 47.03 & \cellcolor{gray!18}\textbf{50.99} \\
5\% & 50.85 & 53.76 & \cellcolor{gray!18}\textbf{60.06} \\
10\% & 54.07 & 57.80 & \cellcolor{gray!18}\textbf{63.62} \\
20\% & 58.33 & 62.34 & \cellcolor{gray!18}\textbf{66.21} \\
30\% & 61.12 & 64.51 & \cellcolor{gray!18}\textbf{66.56} \\
\bottomrule
\end{tabular}
\end{table}

\begin{table}[H]
\centering
\caption{WIKI TRACE: object hit rate (\%) for canonical \SIEVE{}, MT-SIEVE (Mean-Threshold \SIEVE{}), and DT-SIEVE (Dual-Threshold \SIEVE{}). Best per row in bold.}
\label{tab:app-sieve-wiki_2016u}
\small
\begin{tabular}{lrrr}
\toprule
Cache & Canonical & MT-SIEVE & DT-SIEVE \\
\midrule
0.01\% & \cellcolor{gray!18}\textbf{49.02} & 45.81 & 46.06 \\
0.1\% & 66.98 & 66.36 & \cellcolor{gray!18}\textbf{67.24} \\
1\% & 84.94 & 87.74 & \cellcolor{gray!18}\textbf{90.06} \\
5\% & 93.15 & 95.37 & \cellcolor{gray!18}\textbf{96.80} \\
10\% & 95.52 & 97.08 & \cellcolor{gray!18}\textbf{97.80} \\
20\% & 97.19 & 98.09 & \cellcolor{gray!18}\textbf{98.33} \\
30\% & 97.87 & 98.42 & \cellcolor{gray!18}\textbf{98.48} \\
\bottomrule
\end{tabular}
\end{table}

\subsection{Runtime and Byte Hit Rate}
\label{app:sieve-runtime}

\begin{figure}[H]
\centering
\includegraphics[width=\linewidth]{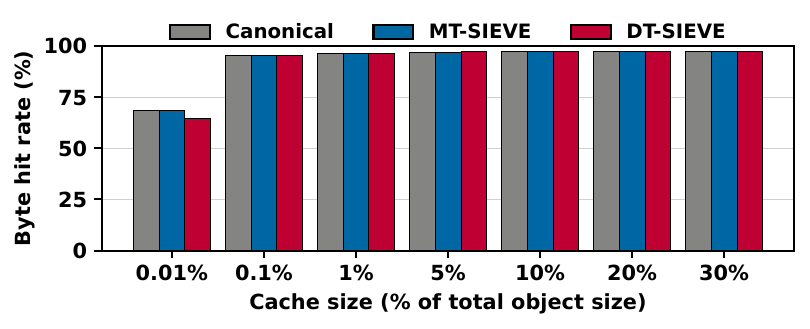}
\caption{Byte hit rate on META TRACE 1 for canonical \SIEVE{}, MT-SIEVE
(Mean-Threshold \SIEVE{}), and DT-SIEVE (Dual-Threshold \SIEVE{}). Byte hit rate
is the fraction of requested bytes served from cache.}
\Description{Grouped bar chart of byte hit rate versus cache size for canonical SIEVE, MT-SIEVE, and DT-SIEVE on META TRACE 1.}
\label{fig:app-sieve-bhr}
\end{figure}

This subsection supplements Section~\ref{sec:sieve-runtime}.
Figure~\ref{fig:app-sieve-bhr} gives the byte hit rate of the three \SIEVE{}
policies on META TRACE 1, and Table~\ref{tab:app-sieve-timing} reports their
whole-trace policy time on the three META traces at the $10\%$ and $30\%$ cache
sizes.

\begin{table}[H]
\centering
\caption{Whole-trace policy time for canonical \SIEVE{}, MT-SIEVE
(Mean-Threshold \SIEVE{}), and DT-SIEVE (Dual-Threshold \SIEVE{}) on the META
traces, in nanoseconds per request. The $10\%$ rows are
the values plotted in Figure~\ref{fig:sieve-wall-clock}. Lower is better. The
fastest policy in each row is bold.}
\label{tab:app-sieve-timing}
\small
\begin{tabular}{llrrr}
\toprule
Trace & Cache & Canonical & MT-SIEVE & DT-SIEVE \\
& & (ns/req) & (ns/req) & (ns/req) \\
\midrule
META TRACE 1 & 10\% & \textbf{122.3} & 233.4 & 195.1 \\
 & 30\% & \textbf{128.5} & 209.8 & 156.8 \\
\midrule
META TRACE 2 & 10\% & \textbf{208.5} & 381.9 & 276.0 \\
 & 30\% & \textbf{210.4} & 288.4 & 225.2 \\
\midrule
META TRACE 3 & 10\% & \textbf{192.4} & 329.0 & 270.2 \\
 & 30\% & \textbf{170.0} & 266.6 & 220.2 \\
\bottomrule
\end{tabular}

\end{table}

\subsection{Synthetic Traces}
\label{app:sieve-synthetic}

This subsection supplements Section~\ref{sec:sieve-synthetic}.
Table~\ref{tab:app-sieve-synthetic} lists the synthetic-trace results plotted in
Figure~\ref{fig:sieve-synthetic}, with the relative gain of each size-aware
variant over canonical \SIEVE{}.

\begin{table}[H]
\centering
\caption{Synthetic-trace object hit rate (\%) for canonical \SIEVE{}, MT-SIEVE
(Mean-Threshold \SIEVE{}), and DT-SIEVE (Dual-Threshold \SIEVE{}): the values
plotted in Figure~\ref{fig:sieve-synthetic}. Gain columns report the relative
improvement over canonical \SIEVE{} at the same trace and cache size. The best
object hit rate in each row is bold and shaded.}
\label{tab:app-sieve-synthetic}
\scriptsize
\setlength{\tabcolsep}{3pt}
\resizebox{\linewidth}{!}{%
\begin{tabular}{llrrrrr}
\toprule
Trace & Cache & Canonical OHR & MT-SIEVE OHR & MT-SIEVE gain & DT-SIEVE OHR & DT-SIEVE gain \\
\midrule
$U(1,100)$ & 0.1\% & \cellcolor{gray!18}\textbf{4.59} & 4.58 & -0.1\% & 4.58 & -0.1\% \\
 & 1\% & 11.76 & 11.77 & +0.1\% & \cellcolor{gray!18}\textbf{11.78} & +0.1\% \\
 & 5\% & 23.45 & \cellcolor{gray!18}\textbf{24.06} & +2.6\% & 24.04 & +2.5\% \\
 & 10\% & 32.14 & \cellcolor{gray!18}\textbf{35.14} & +9.3\% & 35.11 & +9.2\% \\
 & 20\% & 44.29 & 49.95 & +12.8\% & \cellcolor{gray!18}\textbf{49.96} & +12.8\% \\
 & 30\% & 53.61 & \cellcolor{gray!18}\textbf{60.35} & +12.6\% & 60.33 & +12.6\% \\
\midrule
$U(1,500)$ & 0.1\% & \cellcolor{gray!18}\textbf{4.59} & 4.58 & -0.1\% & 4.58 & -0.1\% \\
 & 1\% & 11.75 & 11.77 & +0.2\% & \cellcolor{gray!18}\textbf{11.78} & +0.3\% \\
 & 5\% & 23.45 & 23.98 & +2.3\% & \cellcolor{gray!18}\textbf{23.98} & +2.3\% \\
 & 10\% & 32.14 & 35.06 & +9.1\% & \cellcolor{gray!18}\textbf{35.06} & +9.1\% \\
 & 20\% & 44.29 & \cellcolor{gray!18}\textbf{49.89} & +12.6\% & 49.89 & +12.6\% \\
 & 30\% & 53.61 & 60.29 & +12.5\% & \cellcolor{gray!18}\textbf{60.30} & +12.5\% \\
\midrule
LN(CV=2) & 0.1\% & 4.25 & \cellcolor{gray!18}\textbf{4.26} & +0.1\% & 4.25 & -0.1\% \\
 & 1\% & 12.48 & 12.49 & +0.1\% & \cellcolor{gray!18}\textbf{12.51} & +0.3\% \\
 & 5\% & 23.72 & 27.51 & +16.0\% & \cellcolor{gray!18}\textbf{28.22} & +18.9\% \\
 & 10\% & 32.41 & 39.93 & +23.2\% & \cellcolor{gray!18}\textbf{41.83} & +29.1\% \\
 & 20\% & 44.30 & 56.24 & +27.0\% & \cellcolor{gray!18}\textbf{60.60} & +36.8\% \\
 & 30\% & 53.64 & 67.09 & +25.1\% & \cellcolor{gray!18}\textbf{73.31} & +36.7\% \\
\midrule
LN(CV=4) & 0.1\% & \cellcolor{gray!18}\textbf{3.41} & 3.40 & -0.4\% & 3.41 & -0.1\% \\
 & 1\% & 12.86 & 12.89 & +0.2\% & \cellcolor{gray!18}\textbf{12.91} & +0.4\% \\
 & 5\% & 24.01 & 31.54 & +31.4\% & \cellcolor{gray!18}\textbf{33.74} & +40.5\% \\
 & 10\% & 32.29 & 45.10 & +39.7\% & \cellcolor{gray!18}\textbf{49.94} & +54.7\% \\
 & 20\% & 44.23 & 62.10 & +40.4\% & \cellcolor{gray!18}\textbf{72.08} & +62.9\% \\
 & 30\% & 53.58 & 72.50 & +35.3\% & \cellcolor{gray!18}\textbf{79.85} & +49.0\% \\
\bottomrule
\end{tabular}
}
\end{table}

\subsection{More Retention Tokens}
\label{app:sieve-k8}

This subsection supplements the discussion of retention tokens in
Section~\ref{sec:discussion}. Table~\ref{tab:app-sieve-k8} reports the results for
\SIEVE{}-K8.

\begin{table}[H]
\centering
\caption{Larger-token \SIEVE{} comparison at the $10\%$ cache size.
\SIEVE{}-K8 uses up to eight retention tokens per resident object. The table
shows that additional tokens can improve object hit rate, especially on the
synthetic traces; the main paper reports the two-token rules. OHR columns are object hit rates in
percent; gain columns are relative improvements.}
\label{tab:app-sieve-k8}
\small
\setlength{\tabcolsep}{4pt}
\resizebox{\textwidth}{!}{%
\begin{tabular}{lrrrrr}
\toprule
Trace & Canonical & Dual-Thr. & SIEVE-K8 & K8 gain over canonical & K8 gain over dual \\
\midrule
META TRACE 1 & 66.86 & 71.87 & \cellcolor{gray!18}\textbf{72.44} & +8.3\% & +0.8\% \\
META TRACE 2 & 48.52 & 60.57 & \cellcolor{gray!18}\textbf{61.64} & +27.0\% & +1.8\% \\
META TRACE 3 & 54.07 & 63.62 & \cellcolor{gray!18}\textbf{65.38} & +20.9\% & +2.8\% \\
WIKI TRACE & 95.52 & 97.80 & \cellcolor{gray!18}\textbf{97.83} & +2.4\% & +0.0\% \\
$U(1,100)$ & 32.14 & 35.11 & \cellcolor{gray!18}\textbf{40.92} & +27.3\% & +16.5\% \\
$U(1,500)$ & 32.14 & 35.06 & \cellcolor{gray!18}\textbf{41.12} & +27.9\% & +17.3\% \\
LN(CV=2) & 32.41 & 41.83 & \cellcolor{gray!18}\textbf{51.73} & +59.6\% & +23.7\% \\
LN(CV=4) & 32.29 & 49.94 & \cellcolor{gray!18}\textbf{62.84} & +94.6\% & +25.8\% \\
\bottomrule
\end{tabular}
}
\end{table}

\section{Full Overall Policy Results}
\label{app:overall-results}

This appendix supplements Section~\ref{sec:overall-results}; its subsections
follow the subsections of that section.

\subsection{Synthetic Traces}
\label{app:overall-synthetic}

This subsection supplements Section~\ref{sec:overall-synthetic}.
Table~\ref{tab:app-section5-synthetic-ohr-full} gives the full measured
object-hit-rate sweep for the policies of Section~\ref{sec:overall-results} on the
four synthetic traces at $\alpha=0.6$.

\begin{table}[H]
\centering
\caption{Full measured synthetic object-hit-rate sweep at $\alpha=0.6$ for the Section~\ref{sec:overall-results} policies, which include the existing size-aware policies (LRU\textsuperscript{!}, i.e., LRU-MIN, LHD, and AdaptSize).}
\label{tab:app-section5-synthetic-ohr-full}
\scriptsize
\setlength{\tabcolsep}{1.8pt}
\resizebox{\textwidth}{!}{%
\begin{tabular}{llrrrrrrrrrrr}
\toprule
Trace & Cache & LRU\textsuperscript{!} & AdaptSize & LHD & FIFO & SIEVE\textsuperscript{!} & 2Q\textsuperscript{!} & 2Q-SIEVE\textsuperscript{!} & ARC\textsuperscript{!} & ARC-SIEVE\textsuperscript{!} & S3-FIFO\textsuperscript{!} & S3-FIFO-SIEVE\textsuperscript{!} \\
\midrule
$U(1,100)$ & 0.01\% & 0.7 & 0.9 & 0.7 & 0.1 & 0.2 & 0.7 & 0.4 & 0.7 & 0.4 & \cellcolor{gray!18}\textbf{2.1} & 0.4 \\
 & 0.1\% & 2.9 & 4.0 & 3.0 & 1.0 & 4.6 & 2.8 & 2.5 & 2.9 & 2.7 & \cellcolor{gray!18}\textbf{5.4} & 2.6 \\
 & 1\% & 10.1 & 13.0 & 11.4 & 5.2 & 11.8 & 9.6 & 11.8 & 10.1 & 12.3 & \cellcolor{gray!18}\textbf{13.4} & 11.8 \\
 & 2\% & 14.7 & 17.6 & 16.4 & 8.2 & 14.1 & 14.3 & 17.7 & 14.7 & \cellcolor{gray!18}\textbf{18.4} & 17.6 & 17.7 \\
 & 5\% & 23.6 & 26.8 & 25.9 & 14.9 & 24.0 & 23.2 & 28.8 & 23.5 & \cellcolor{gray!18}\textbf{29.7} & 25.3 & 28.8 \\
 & 10\% & 31.9 & 36.5 & 35.8 & 23.1 & 35.1 & 31.6 & 40.0 & 31.8 & \cellcolor{gray!18}\textbf{40.9} & 33.6 & 40.1 \\
 & 20\% & 44.0 & 47.7 & 48.8 & 35.8 & 50.0 & 43.7 & 53.1 & 43.9 & \cellcolor{gray!18}\textbf{53.6} & 44.8 & 53.1 \\
 & 30\% & 53.4 & 54.5 & 58.1 & 45.9 & 60.3 & 53.3 & 61.2 & 53.4 & \cellcolor{gray!18}\textbf{61.5} & 53.3 & 61.3 \\
$U(1,500)$ & 0.01\% & 1.0 & 1.1 & 0.7 & 0.1 & 0.2 & 0.9 & 0.4 & 1.0 & 0.4 & \cellcolor{gray!18}\textbf{2.1} & 0.4 \\
 & 0.1\% & 3.3 & 4.2 & 3.0 & 1.0 & 4.6 & 3.1 & 2.6 & 3.3 & 2.8 & \cellcolor{gray!18}\textbf{5.4} & 2.6 \\
 & 1\% & 10.2 & 13.2 & 11.5 & 5.2 & 11.8 & 9.8 & 11.9 & 10.2 & 12.5 & \cellcolor{gray!18}\textbf{13.4} & 11.9 \\
 & 2\% & 14.8 & 17.7 & 16.6 & 8.2 & 14.1 & 14.5 & 17.9 & 14.8 & \cellcolor{gray!18}\textbf{18.6} & 17.6 & 17.9 \\
 & 5\% & 24.8 & 27.0 & 26.0 & 14.9 & 24.0 & 24.0 & 29.1 & 24.7 & \cellcolor{gray!18}\textbf{30.0} & 25.3 & 29.1 \\
 & 10\% & 33.7 & 36.6 & 36.0 & 23.1 & 35.1 & 33.3 & 40.3 & 33.6 & \cellcolor{gray!18}\textbf{41.1} & 33.6 & 40.3 \\
 & 20\% & 45.6 & 47.8 & 49.0 & 35.8 & 49.9 & 45.3 & 53.2 & 45.5 & \cellcolor{gray!18}\textbf{53.7} & 44.8 & 53.3 \\
 & 30\% & 54.7 & 54.6 & 58.2 & 45.9 & 60.3 & 54.5 & 61.3 & 54.7 & \cellcolor{gray!18}\textbf{61.7} & 53.3 & 61.4 \\
LN(CV=2) & 0.01\% & 0.9 & \cellcolor{gray!18}\textbf{1.6} & 1.0 & 0.1 & 0.3 & 0.8 & 0.5 & 0.9 & 0.6 & 0.2 & 0.5 \\
 & 0.1\% & 4.4 & 5.5 & 4.2 & 1.0 & 4.2 & 4.2 & 3.1 & 4.4 & 3.3 & \cellcolor{gray!18}\textbf{5.8} & 3.1 \\
 & 1\% & 15.9 & \cellcolor{gray!18}\textbf{17.1} & 16.1 & 5.2 & 12.5 & 15.0 & 13.9 & 15.9 & 14.8 & 15.1 & 13.9 \\
 & 2\% & 22.9 & \cellcolor{gray!18}\textbf{23.9} & 23.3 & 8.2 & 16.3 & 21.6 & 21.2 & 22.8 & 22.5 & 19.7 & 21.2 \\
 & 5\% & 35.8 & 35.6 & 36.0 & 14.9 & 28.2 & 34.4 & 35.5 & 36.0 & \cellcolor{gray!18}\textbf{37.3} & 27.9 & 35.5 \\
 & 10\% & 48.1 & 46.6 & 48.5 & 23.1 & 41.8 & 47.2 & 49.8 & 49.0 & \cellcolor{gray!18}\textbf{51.9} & 36.6 & 49.9 \\
 & 20\% & 60.7 & 57.0 & 62.8 & 35.7 & 60.6 & 61.7 & 65.0 & 63.8 & \cellcolor{gray!18}\textbf{66.7} & 48.3 & 65.1 \\
 & 30\% & 67.2 & 60.4 & 71.2 & 45.9 & 73.3 & 70.7 & 72.4 & 72.5 & \cellcolor{gray!18}\textbf{73.8} & 57.1 & 72.6 \\
LN(CV=4) & 0.01\% & 1.1 & \cellcolor{gray!18}\textbf{2.2} & 1.1 & 0.1 & 0.4 & 0.9 & 0.6 & 1.1 & 0.7 & 0.1 & 0.6 \\
 & 0.1\% & 6.7 & \cellcolor{gray!18}\textbf{8.0} & 5.2 & 1.0 & 3.4 & 6.1 & 4.0 & 6.7 & 4.3 & 3.2 & 4.0 \\
 & 1\% & \cellcolor{gray!18}\textbf{25.3} & 23.5 & 19.8 & 5.2 & 12.9 & 24.1 & 18.2 & 25.3 & 19.3 & 16.1 & 18.2 \\
 & 2\% & \cellcolor{gray!18}\textbf{34.5} & 30.6 & 28.5 & 8.2 & 19.6 & 32.8 & 27.4 & 34.4 & 29.0 & 20.8 & 27.4 \\
 & 5\% & 49.0 & 43.6 & 43.4 & 14.9 & 33.7 & 47.2 & 44.9 & \cellcolor{gray!18}\textbf{49.2} & 47.2 & 29.3 & 44.9 \\
 & 10\% & 60.7 & 54.7 & 56.4 & 23.1 & 49.9 & 59.8 & 61.0 & 61.7 & \cellcolor{gray!18}\textbf{63.2} & 38.1 & 61.0 \\
 & 20\% & 71.4 & 63.8 & 69.5 & 35.7 & 72.1 & 72.5 & 73.8 & 74.2 & \cellcolor{gray!18}\textbf{74.9} & 50.2 & 73.9 \\
 & 30\% & 76.2 & 63.8 & 76.3 & 45.9 & 79.8 & 79.1 & 78.8 & \cellcolor{gray!18}\textbf{80.4} & 79.6 & 59.1 & 78.9 \\
\bottomrule
\end{tabular}%
}
\end{table}

\subsection{Runtime Overhead and Byte Hit Rate}
\label{app:overall-runtime}

\begin{figure}[H]
\centering
\includegraphics[width=\textwidth]{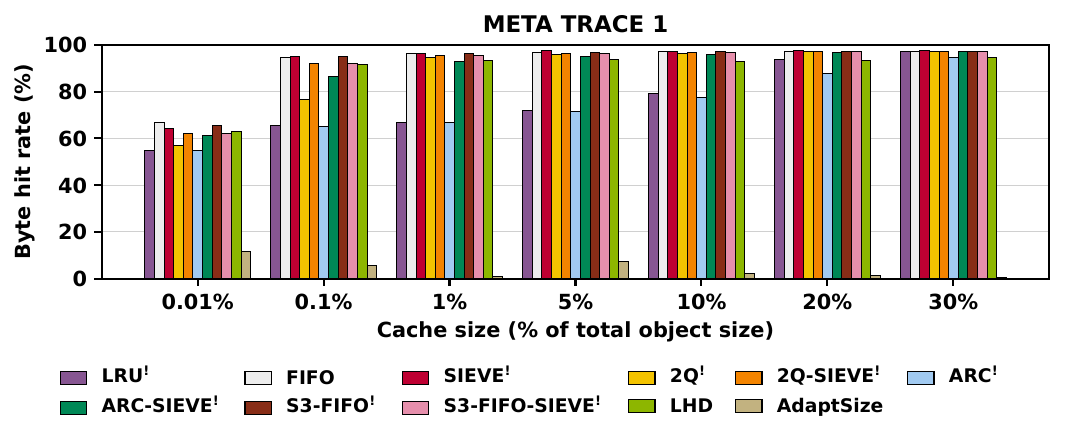}
\caption{Byte hit rate on META TRACE 1 for our proposed policies and
the existing size-aware policies (LRU\textsuperscript{!}, LHD, and AdaptSize). Byte hit rate is the
fraction of requested bytes served from cache. LRU\textsuperscript{!} is LRU-MIN
and \SIEVE{}\textsuperscript{!} is Dual-Threshold \SIEVE{}.}
\Description{Grouped bar chart of byte hit rate versus cache size for eleven policies on META TRACE 1.}
\label{fig:app-section5-bhr-meta1}
\end{figure}

This subsection supplements Section~\ref{sec:overall-runtime}.
Table~\ref{tab:app-section5-timing-meta} gives the policy time on the META traces,
and Figure~\ref{fig:app-section5-bhr-meta1} shows the byte hit rate on META
TRACE~1.

\begin{table}[H]
\centering
\caption{Whole-trace policy time on the META traces, in nanoseconds per request. Lower is better and the fastest policy in each row is bold. FIFO is libCacheSim's FIFO. Values marked $^\dagger$ are the elapsed time of the whole FIFO replay, including trace decoding, divided by the number of requests, and are therefore upper bounds on FIFO's policy-only time; the unmarked FIFO values are policy-only times measured in the same way as the other policies, from a single run.}
\label{tab:app-section5-timing-meta}
\scriptsize
\setlength{\tabcolsep}{3pt}
\resizebox{\textwidth}{!}{%
\begin{tabular}{llrrrrrrrrr}
\toprule
Trace & Cache & LRU\textsuperscript{!} & FIFO & SIEVE\textsuperscript{!} & 2Q\textsuperscript{!} & 2Q-SIEVE\textsuperscript{!} & ARC\textsuperscript{!} & ARC-SIEVE\textsuperscript{!} & S3-FIFO\textsuperscript{!} & S3-FIFO-SIEVE\textsuperscript{!} \\
\midrule
META TRACE 1 & 10\% & 1243 & \textbf{50} & 195 & 1222 & 195 & 1492 & 349 & 321 & 202 \\
META TRACE 1 & 30\% & 1326 & \textbf{114}$^\dagger$ & 157 & 1116 & 188 & 1354 & 288 & 308 & 195 \\
META TRACE 2 & 10\% & 1925 & \textbf{83} & 276 & 1916 & 300 & 2357 & 509 & 539 & 314 \\
META TRACE 2 & 30\% & 2129 & 249$^\dagger$ & \textbf{225} & 1849 & 273 & 2140 & 398 & 448 & 284 \\
META TRACE 3 & 10\% & 1744 & \textbf{228}$^\dagger$ & 270 & 1725 & 265 & 2200 & 518 & 472 & 277 \\
META TRACE 3 & 30\% & 1910 & \textbf{175}$^\dagger$ & 220 & 1665 & 276 & 1969 & 406 & 413 & 278 \\
\bottomrule
\end{tabular}%
}
\end{table}

\section{\SIEVE{} Working-Set Approximation Details}
\label{app:wsa-details}

This appendix gives the full working-set approximation (WSA) derivation and validation. Figure~\ref{fig:app-wsa-metapop} compares the WSA predictions with empirical \SIEVE{} replays for the measured popularities of META TRACE 1, and Figures~\ref{fig:app-wsa-unit} and~\ref{fig:app-wsa-var} compare them for Zipf-popularity unit-size and variable-length objects at $N=10^6$ and $N=10^7$.

\SIEVE{} keeps (equal-sized) objects in FIFO insertion order and uses one visited bit per object together with an eviction hand. On a cache hit, the object's visited bit is set to one; the object is not moved and the hand does not advance. On a miss, if an eviction is needed, the hand examines objects in order of a circular cache. A visited object is kept in place and if its bit is set to one then its bit is cleared to zero and the hand continues. The first object whose bit is  zero is evicted. The missed object is then inserted at the FIFO head with visited bit zero. Thus a newly inserted object can be evicted the first time the hand reaches it if it has not been hit. \SIEVE{} also defines a logical examination clock and shows that inter-examination times depend on queue position~\cite{ZYY24}. 
This suggests that a WSA for \SIEVE{} should be based on hand examinations rather than on LRU recency alone.

\subsection{\SIEVE{} Working-Set Approximation with Two Examination Times}
Consider a cache that holds $B$ equal-sized objects. There are $N$ objects, and each object $i$ is requested according to an independent Poisson process of rate $\lambda_i>0$, i.e., the Independent Reference Model (IRM). Let
\begin{equation}
\Lambda=\sum_{i=1}^{N}\lambda_i,
\qquad
p_i=\frac{\lambda_i}{\Lambda},
\qquad
\sum_{i=1}^{N}p_i=1.
\end{equation}
For comparison, the usual LRU characteristic-time approximation
chooses (constant) $T_C$ from
\begin{equation}
\sum_{i=1}^{N}\left(1-e^{-\lambda_i T_C}\right)=B,
\end{equation}
and then uses $1-e^{-\lambda_iT_C}$ as the occupancy probability of object $i$. By Poisson Arrivals See Time Averages (PASTA)~\cite{WolffBook89}, this occupancy probability is also the hit probability under the IRM~\cite{DeS72,CTZ02}.

For \SIEVE{}, the first hand examination of a newly admitted object is different from a subsequent examination of an object that has already survived once. We therefore use two characteristic examination times:
\begin{align*}
A & =\text{characteristic time from admission to first examination}\\
T & =\text{characteristic time between subsequent examinations}
\end{align*}

Suppose object $i$ has already survived one hand examination. At that
examination, its visited bit is reset to $0$. We define the constant $T$ as the
approximate time until the \SIEVE{} hand returns to and examines the same
object again.
Actual \SIEVE{} examination times are random and depend on the hand position.

Recall that a newly admitted object starts with visited bit zero, and it survives its first examination only if it receives at least one request during the interval of length $A$, so we approximate
\begin{equation}
 a_i(A)=\Pr\{\text{survive first examination}\}
       =1-e^{-\lambda_i A}.
\end{equation}
When an object survives an examination, its visited bit is cleared to zero.
During the subsequent interval of length $T$, the probability of at least one request is
\begin{equation}
 s_i(T)=1-e^{-\lambda_iT};
\end{equation}
and the probability of no request, so that the object is evicted at the next examination, is
\begin{equation}
 q_i(T)=e^{-\lambda_iT}.
\end{equation}

Let $G_i$ be the number of subsequent $T$-intervals, including the final interval in which no request occurs. Under this approximation, $G_i$ is geometrically distributed,
\begin{equation}
\Pr\{G_i=k\}=s_i(T)^{k-1}q_i(T), \qquad k\ge 1,
\end{equation}
with mean
\begin{equation}
\mathbb{E}[G_i]
=\frac{1}{q_i(T)}
=e^{\lambda_iT}.
\end{equation}
After an object has survived once, $G_i$ denotes the number of
additional examination intervals that the object remains in the cache
before an interval occurs in which it receives no hit.

Every admission keeps the object in the cache for the first interval $A$. With probability $a_i(A)$ the object survives that first examination and then remains for an expected $\mathbb{E}[G_i]$ subsequent intervals. The mean residence time after admission is therefore
\begin{align}
R_i(A,T)
&=A+a_i(A)T\mathbb{E}[G_i] \\
&=A+\left(1-e^{-\lambda_iA}\right)T e^{\lambda_iT}.
\end{align}

When object $i$ is not in the cache, the next request for it arrives
after an average time $1/\lambda_i$. That request is a miss and brings
the object back into the cache. The object then stays in the cache for
an average time $R_i(A,T)$. Therefore, the long-run fraction of time
that object $i$ is in the cache is
\begin{align}
h_i(A,T)
~=~\frac{R_i(A,T)}{R_i(A,T)+1/\lambda_i} 
~=~\frac{\lambda_iR_i(A,T)}{1+\lambda_iR_i(A,T)}.
\end{align}
By the PASTA property, under the IRM, $h_i(A,T)$ is also the hit probability of object $i$. The cache-capacity condition is therefore
\begin{equation}
\sum_{i=1}^{N} h_i(A,T)=B.
\end{equation}
The aggregate hit and miss ratios are
\begin{equation}
H(A,T)=\sum_{i=1}^{N}p_i h_i(A,T),
\qquad
M(A,T)=1-H(A,T).
\end{equation}
At this point there is one capacity equation but two characteristic times, $A$ and $T$. The remaining task is to relate them using the \SIEVE{} hand process.

\subsection{A Method to Relate the First and Subsequent Examination Times}

 \SIEVE{} measures inter-examination time using the
number of objects examined by the hand. We use
this examination-based view to relate $A$ and $T$.

Let $K_0$ be the number of hand examinations from the admission of a typical object until its first examination. Let $K_1$ be the number of hand examinations between two subsequent examinations of an object that has already survived. Let $\nu$ denote the long-run mean examination rate, in examinations per unit time. A first mean-field conversion is
\begin{equation}
A\approx \frac{\mathbb{E}[K_0]}{\nu},
\qquad
T\approx \frac{\mathbb{E}[K_1]}{\nu}.
\end{equation}
The examination rate $\nu$ already contains the two features that make SIEVE different from LRU: only misses start an eviction scan, and one miss may examine several objects before finding a zero bit. Taking the ratio removes this common rate:
\begin{equation}
\rho:=\frac{A}{T}
\approx
\frac{\mathbb{E}[K_0]}{\mathbb{E}[K_1]}.
\end{equation}
Thus a practical closure is
\begin{equation}
A=\rho T.
\end{equation}
For a chosen value of $\rho$, the residence time becomes
\begin{align}
R_i(T;\rho)
&=\rho T+
\left(1-e^{-\rho\lambda_iT}\right)T e^{\lambda_iT},
\end{align}
and
\begin{equation}
h_i(T;\rho)
=\frac{\lambda_iR_i(T;\rho)}{1+\lambda_iR_i(T;\rho)}.
\end{equation}
The characteristic time $T$ is then obtained from the single equation
\begin{equation}
\sum_{i=1}^{N}h_i(T;\rho)=B.
\end{equation}
For fixed $\rho>0$, this equation can be solved numerically by bisection. Once $T$ is known, the hit and miss ratios follow from the same $h_i$ values.
For a simple baseline for $\rho$, suppose that a new admission occurs at a uniformly random point during a representative \SIEVE{} hand round. Let $K$ be the number of object examinations in one representative round. Under the uniform-phase assumption, the expected number of examinations from admission until the end of the round is approximately $K/2$, so $\mathbb{E}[K_0] \approx K/2$. For an object that has already survived an examination, we approximate the time until its next examination by one full round, so $\mathbb{E}[K_1] \approx K$. Therefore,
\[
\rho
\approx
\frac{\mathbb{E}[K_0]}{\mathbb{E}[K_1]}
\approx
\frac{K/2}{K}
=
\frac{1}{2}.
\]
Again, this value $\rho \approx 1/2$ is only a baseline resulting from the
uniform-phase assumption; it is not a property of \SIEVE{}. 
\cite{ZYY24} shows that \SIEVE{} inter-examination times depend on queue
position.

\subsection{Variable-Length Objects for \SIEVE{}$'$}

For variable-length objects, let $L_i$ denote the size of object $i$ in bytes. To keep the notation consistent with the unit-size formulation, we continue to use $B$ for cache capacity; in this section, however, $B$ denotes cache capacity in bytes, i.e., the byte budget, rather than a number of object slots. The approximation for $R_i(T;\rho)$ and $h_i(T;\rho)$ is unchanged. Only the capacity equation is replaced by the byte-weighted condition
\begin{equation}
\sum_{i=1}^{N} L_i h_i(T;\rho)=B.
\label{eq:wsa-byte-cap}
\end{equation}
For fixed $\rho$, this equation determines $T$ numerically. If $L_i>B$, object $i$ cannot be admitted and its hit probability is set to zero. The aggregate hit ratio remains
\begin{equation}
H(T;\rho)=\sum_{i=1}^{N} p_i h_i(T;\rho).
\label{eq:wsa-var-hit}
\end{equation} 
Thus, the variable-length form changes the capacity accounting while retaining the same per-object working-set approximation.

\subsection{Evaluation of \SIEVE{} and \SIEVE{}$'$ WSA with META Popularities}
\label{app:wsa-meta-pop}

\begin{figure}[H]
\centering
\begin{minipage}[t]{0.485\linewidth}
\centering
\includegraphics[width=\linewidth]{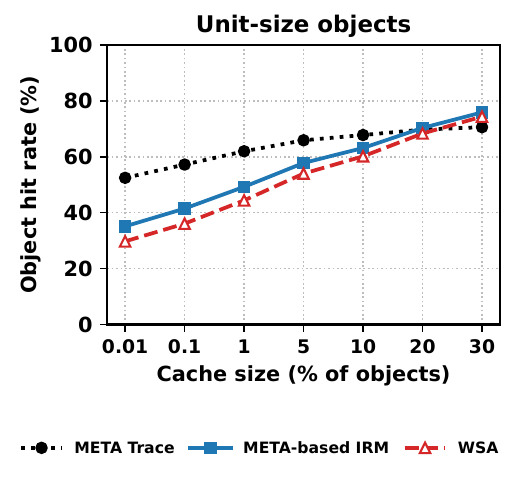}
{\small (a) unit-size objects}
\end{minipage}\hfill
\begin{minipage}[t]{0.485\linewidth}
\centering
\includegraphics[width=\linewidth]{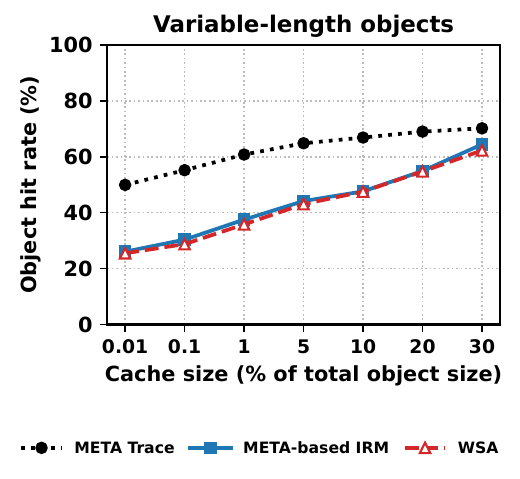}
{\small (b) variable-length objects}
\end{minipage}
\caption{Object hit rate for META TRACE 1 with (a) \SIEVE{} for unit-size objects
and (b) \SIEVE{}$'$ for variable-length objects. META Trace: the trace in its original request order and times.
META-based IRM: independent Poisson requests based on the popularities from the trace.
WSA: the working-set approximation for the IRM with the same popularities. The cache size is
a percentage of the number of objects in (a) and of the total object size in (b).}
\Description{Two line charts, for unit-size and variable-length objects, each comparing the SIEVE hit rate on the META trace, on an IRM stream with the popularities of the trace, and the working-set-approximation prediction, versus cache size.}
\label{fig:app-wsa-metapop}
\end{figure}

We first validate the WSA using the object popularities of META TRACE 1. Let
$n_i$ be the number of requests for object $i$ and $Q$ the total number of
requests in the trace, and set $p_i=n_i/Q$; each object retains its size $L_i$.
Figures~\ref{fig:app-wsa-metapop}(a) and~\ref{fig:app-wsa-metapop}(b) show three
hit-rate curves for \SIEVE{} and \SIEVE{}$'$, respectively. \emph{META Trace} is
the trace replayed in its original request order and times. \emph{META-based IRM} is a long
request stream in which every request is drawn independently as a Poisson process with probabilities
$p_i$, and its hit rate is measured in steady state. This stream has the same
popularities and sizes as the trace, but not the same request order. \emph{WSA}
is the hit rate predicted by the working-set approximation of the IRM for the same $p_i$.
For unit-size objects, every object has size one and $B$ is a number of objects. 
For variable-length objects, $B$ is in bytes and the capacity equation is
\eqref{eq:wsa-byte-cap}. Since the WSA assumes the IRM, it should be compared
with the META-based IRM curve. The two are close in both cases and nearly
coincide for variable-length objects.



\subsection{Evaluation of \SIEVE{} WSA for Equal-Size Objects with Zipf Popularities}
\label{app:wsa-unit}

We next validate the working-set approximation using synthetic IRM request
streams with Zipf popularity,
\begin{equation}
p_i =
\frac{i^{-\alpha}}
{\sum_{j=1}^{N} j^{-\alpha}},
\qquad i=1,\ldots,N.
\end{equation}
All objects have unit size. Therefore, $B$ denotes the number of
objects that can be stored in the cache. We consider
$N\in\{10^6,10^7\}$ and $\alpha\in\{0.8,1.0,1.2\}$.

For each configuration, we compare the working-set-approximation (WSA) prediction
$H_{\mathrm{WSA}}$ with the empirical hit ratio $H_{\mathrm{empirical}}$ measured
by replaying the same request stream through \SIEVE{}.

\begin{figure}[H]
\centering
\begin{minipage}[t]{0.485\linewidth}
\centering
\includegraphics[width=\linewidth]{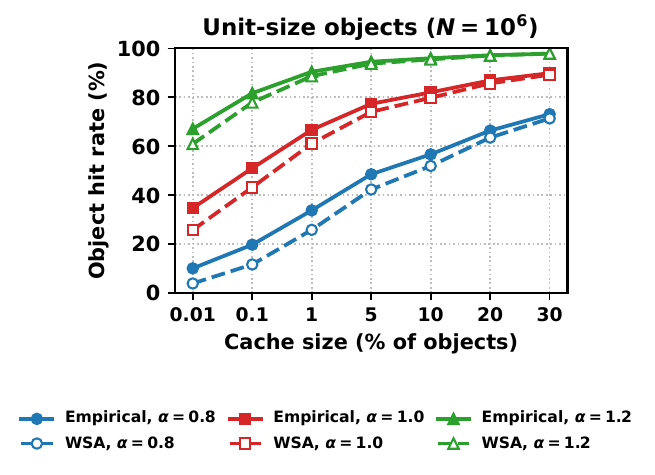}
{\small (a) $N=10^6$}
\end{minipage}\hfill
\begin{minipage}[t]{0.485\linewidth}
\centering
\includegraphics[width=\linewidth]{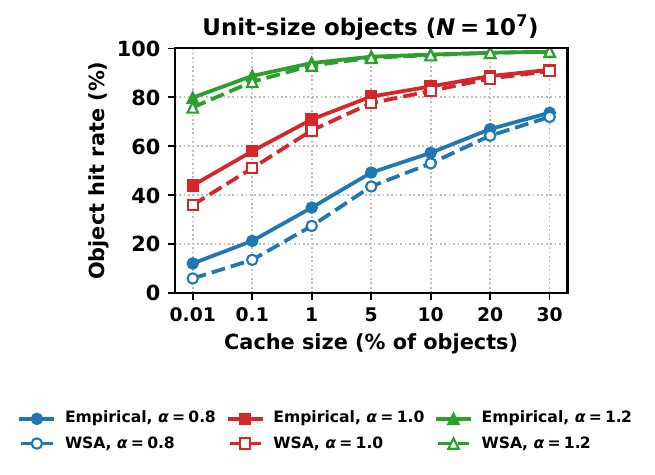}
{\small (b) $N=10^7$}
\end{minipage}
\caption{WSA versus empirical \SIEVE{} hit rate for unit-size objects with (a) $N=10^6$
and (b) $N=10^7$ objects and $Q=100N$ Zipf requests with
$\alpha\in\{0.8,1.0,1.2\}$. Solid lines with filled markers are the empirical
replay; dashed lines with open markers are the WSA prediction. The cache size is a
percentage of the number of objects. Like any such approximation, the WSA is accurate only
asymptotically, for large $N$ or large cache size $B$.}
\Description{Two line charts, for one million and ten million unit-size objects, comparing working-set-approximation predictions with empirical SIEVE hit rates versus cache size for three Zipf parameters.}
\label{fig:app-wsa-unit}
\end{figure}

Figure~\ref{fig:app-wsa-unit} shows that the
unit-size approximation improves as either $N$ or $B$ increases. For small caches,
especially with $\alpha=0.8$, the WSA underestimates the empirical hit rate. For
large caches, and more so for $N=10^7$, the two curves nearly coincide. As with any such
approximation, the working-set approximation should therefore be read as an
asymptotic description, valid for large $N$ or large $B$, rather than as an exact
prediction for small caches.


\subsection{Evaluation of \SIEVE{}$'$ WSA for Variable-Length Objects with Zipf Popularities}
\label{app:wsa-variable}

For the Zipf-popularity variable-length experiments, we use real object sizes from
META TRACE 1 together with a controlled synthetic Zipf request
distribution. The variable-length workload is constructed as follows.
First, for every distinct object in META TRACE 1, we obtain its
object size $L_i$ and its empirical request count. Second, the objects
are sorted in decreasing order of empirical request count, so that each
popularity rank has an associated real META object size. Third, the
empirical popularity values are discarded and rank $i$ is instead
assigned the Zipf probability,
\begin{equation}
p_i =
\frac{i^{-\alpha}}
{\sum_{j=1}^{N}j^{-\alpha}}.
\end{equation}
Finally, the real META object size associated with each rank is retained.
Thus, synthetic object $i$ is represented by the fixed pair
\begin{equation}
(p_i,L_i).
\end{equation}
Every request for object $i$ therefore has the same size $L_i$.
Object sizes are not independently sampled or resampled for different
requests. This preserves the correspondence between popularity rank
and object size observed in META TRACE 1 while allowing the request
popularity to follow a controlled Zipf distribution.


For variable-length objects, $B$ denotes cache capacity in bytes. The capacity equation is \eqref{eq:wsa-byte-cap} and the reported hit ratio remains \eqref{eq:wsa-var-hit}. Here, $B(\%)$ is the cache capacity as a percentage of the total object size in bytes.

\begin{figure}[H]
\centering
\begin{minipage}[t]{0.485\linewidth}
\centering
\includegraphics[width=\linewidth]{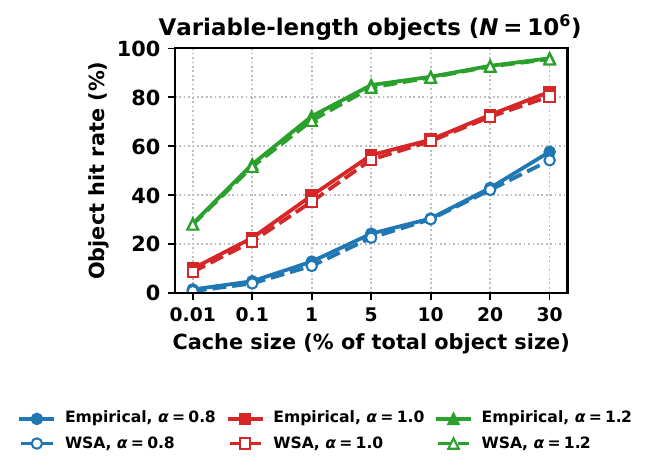}
{\small (a) $N=10^6$}
\end{minipage}\hfill
\begin{minipage}[t]{0.485\linewidth}
\centering
\includegraphics[width=\linewidth]{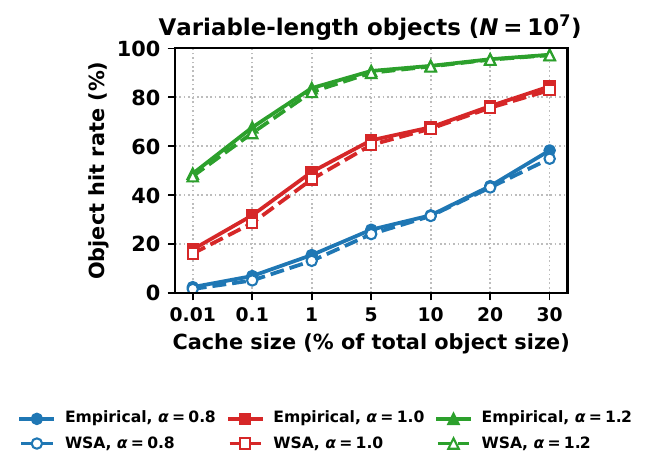}
{\small (b) $N=10^7$}
\end{minipage}
\caption{WSA versus empirical \SIEVE{} hit rate for variable-length objects with (a) $N=10^6$
and (b) $N=10^7$ objects and $Q=100N$ Zipf requests with
$\alpha\in\{0.8,1.0,1.2\}$. Solid lines with filled markers are the empirical
replay; dashed lines with open markers are the WSA prediction. The cache size is a
percentage of the total object size. Like any such approximation, the WSA is accurate only
asymptotically, for large $N$ or large cache size $B$.}
\Description{Two line charts, for one million and ten million variable-length objects, comparing working-set-approximation predictions with empirical SIEVE hit rates versus cache size for three Zipf parameters.}
\label{fig:app-wsa-var}
\end{figure}

Figure~\ref{fig:app-wsa-var} shows that the
variable-length approximation behaves in the same way: it is least accurate for
small caches and most accurate for large $N$ and large $B$, where the WSA and
empirical curves nearly coincide.

\end{document}